\documentclass{aa}  

\usepackage{graphicx}
\usepackage{txfonts}
\begin{document}

   \title{Light breeze in the local Universe}


    \author{A. Concas
          \inst{1}
          \and
          P. Popesso
          \inst{1}
          \and
          M. Brusa
          \inst{2,3}
             \and
	V. Mainieri
		 \inst{4}      
	\and 
	G. Erfanianfar
	\inst{1}
	\and  
	L. Morselli
	\inst{1}          
          }

   \institute{Excellence Cluster Universe,
                Boltzmannstr. 2  D-85748 Garching,  Germany 
         \and
         Dipartimento di Fisica e Astronomia,
              Universit\'a degli Studi di Bologna,
              V.le Berti Pichat, 6/2 - 40127, Bologna, Italy
             \and
             INAF$-$Osservatorio Astronomico di Bologna,
              via Ranzani 1, I-40127, Bologna, Italy
              \and
              European Southern Observatory, 
              Karl-Schwarzschild-str. 2, 85748 Garching, Germany
             }

   \date{Received; accepted}

 
  \abstract
   {We analyze a complete spectroscopic sample of galaxies ($\sim$600,000 ) drawn from Sloan Digital Sky Survey (SDSS, DR7) to look for evidence of galactic winds in the local Universe.
We focus on the shape of the [OIII]$\lambda$5007 emission line as a tracer of ionizing gas outflows. We stack our spectra in a fine grid of  star formation rate (SFR) and stellar mass to analyze the dependence of winds on the position of galaxies in the SFR versus mass diagram. We do not find any significant evidence of broad and shifted [OIII]$\lambda$5007 emission line  which we interpret as no evidence of outflowing ionized gas in the global population.
We have also classified these galaxies as star-forming or AGN dominated according to their position in the standard BPT diagram. We show how the average [OIII]$\lambda$5007 profile changes as function of nature of the dominant ionizing source. We find that in the star-forming dominated source the oxygen line is symmetric and governed by the gravitational potential well. The AGN or composite AGN$\setminus$star-formation activity objects, in contrast, display a prominent and asymmetric profile that can be well described by a broad gaussian component that is blue-shifted from a narrow symmetric core. In particular, we find that the blue wings of the average [OIII]$\lambda$5007 profiles are increasingly prominent in the LINERs and Seyfert galaxies.
We conclude that, in the low-redshift Universe, "pure" star-formation activity does not seem capable of driving ionized-gas outflows, while, the presence of optically selected AGN seems to play a primary role to drive such winds.
We discuss the implications of these results for the role of the quenching mechanism in the present day Universe.}


   \maketitle
%


\section{Introduction}

The most striking feature of the history of our Universe is a drastic decrease in the star formation activity of the galaxy population by almost an order of magnitude over the last 10 Gyr, after a phase of high and rather constant activity \citep[e.g.][for a comprehensive review]{Lilly1996,Madau1998,MadauDickinson2014}. Which process or which combination of processes, so called ``queching" of the star formation activity, causes such decrease is still matter of an intensive debate. 
It is apparent that identifying the quenchin process(es) is crucial for establishing a complete view of how galaxies evolve across cosmic time.

{{According to the most accredited galaxy formation models, from the semi-analytical (SAM) ones to the more recent mass abundance matching models, in the central galaxies, the efficiency in converting the gas fraction in stars reaches a maximum at halo mass $\sim 10^{12}$ M$_{\odot}$ with only $\sim 20$ \% of their baryons currently locked up in stars (see for example \citealp{Croton2006,Guo+11} based on the Millennium Simulation, and \citealp{Moster2010, Behroozi2010, Yang+12} among the mass abundance matching models). The efficiency drops down steeply towards both sides of this mass threshold \citep[e.g.][]{Madau1996,Baldry2008,ConroyWechsler2009,Guo2010,Moster2010,Moster2013,Behroozi2010,Behroozi2013}. 
There is an overall agreement, from the theoretical point of view, that below halo masses of $10^{12}$ M$_{\odot}$, the decreasing SF efficiency is likely to be due to gas eating and removal associated with the star formation activity. Indeed, galactic winds driven by the energy and momentum imprinted by massive stars to the surrounding ISM, are believed to be sufficiently energetic to eject the gas away from the galaxy potential well and quench the star formation (see for instance \citealp{Chevalier77} for energy-driven outflows,  \citealp{Murray+05} for momentum-driven outflows, and to
 \citealp{Hopkins+14} for effect of  multiple stellar feedback in cosmological simulations).
Above stellar masses of $10^{12}$ M$_{\odot}$, instead, more powerful outflows are required to let the gas escape from the deeper galaxy potential well. The energy and radiation generated by accretion onto the massive black hole (BH), in the most massive galaxies, exceeds the binding energy of the gas by a large factor (see  \citealp{Fabian2012} for a complete review). Therefore, energetic feedback from active galactic nuclei (AGN) is believed to provide an important and effective mechanism to eject the gas away by powerful winds, stop the growth of the galaxy and stifle accretion onto the BH  \citep[]{DiMatteo2005,DeLucia2006,Croton2006,Hopkins2006,Bower2006,Hopkins+14, Henriques+16}.}}

{{However, although these models are very successful in reproducing a large variety of observational evidence, in particular, the evolution of the stellar mass function \citep[e.g.][]{Henriques+16}, they still lack a clear observational confirmation. Indeed, a lot of effort has been spent in the last decade from the observational point of view to observe the presence of such outflows at any mass scale and to study their effect on the evolution of the galaxy star formation activity to identify a possible relation of cause and effect. }}
{\cite{Steidel+10} observe blue-shifted Lyman-$\alpha$ emission in most of the SF galaxy population at redshift $\sim$2 and associate such emission line disturbance to SF-induced outflow (see also \citealp{Erb2015} for different emission line study). At somewhat lower redshift, but in a large redshift window ($0.5 < z < 1.5$), \cite{Martin+12} use UV rest-frame absorption features to identify blue-shifted components as indication of outflow and find evidence of outflowing material in massive, highly star forming galaxies (see also \citealp{Rubin+14} for similar studies).  }

{Powerful AGN-driven outflows are recently observed both at low \citep[e.g.][]{Feruglio2010, Villar-Martin2011,RupkeVeilleux2011,Rupke+Veilleux13,Greene2012,Mullaney2013,Rodriguez-Zaurin+13,Cicone2014} and high redshift \citep[e.g.][]{Maiolino2012,Tremonti2007,Brusa2015,Perna2015,Cresci2015}.  However, it is not clear yet if the AGN feedback, in the form of galactic flows, is a specific property of the bulk of the AGN population or if it concerns only a subclass of these objects \cite{Brusa2015}. In addition, it is unclear if they cause a quenching or an enhancement of the galaxy SF activity \citep[e.g.][]{Cresci2015}.}

Furthermore, new observation are revealing the ubiquity of SF induced outflows in very actively star-forming galaxies at all cosmic epochs (see \citealp{Veilleux2005} and  \citealp{Erb2015} for a comprehensive overview). They usually are associated with energetic starburst phenomena \citep[e.g.][]{Heckman+90,Pettini2000,Shapley2003,Rupke2002,Rupke2005a,Rupke2005b,Martin2005,Martin2006, HillZakamska2014}, while their impact in the normal star-forming galaxies \citep{Chen+10,Martin+12,Rubin+14,Cicone2016} and their global effect on the baryon cycle is still debated \cite{Steidel+10}.


These studies have traditionally been carried out with relatively small samples of galaxies. The availability of large spectroscopic data sets such as the SDSS  \cite{York2000}, allows to extend such studies dramatically in size. Significant improvement has been recently made in this regard \citep[e.g.][]{GreeneHo2005a, Chen+10,Mullaney2013,Cicone2016}.
However, all these resent works have focused only on a particular category of galaxies. For example, the optically-selected AGN in \cite{Mullaney2013} and  \cite{GreeneHo2005a} and the star-forming galaxies without AGN contribution in \cite{Chen+10} and \cite{Cicone2016}.

{The aim of this paper is to explore the global proprieties of galactic winds in the local Universe {\it{a)}} by analyzing their incidence in the galaxy population, {\it{b)}} by identifying the powering mechanism: star formation, AGN or a mixed contribution of the two, and {\it{c)}} by clarifying what impact they might have on the galaxy SF activity. For this purpose we investigate the outflow signatures in a large sample of optical spectra at redshift $z<0.3$, drawn from the Sloan Digital Sky Survey (SDSS; \citealp{Abazajian2009}), by using the ionized gas as traced by the [OIII]$\lambda5007$ emission line. The [OIII]$\lambda 5007$ emission line is one of the strongest features in the rest-frame optical 1D spectrum of both active star-forming and AGN dominated galaxies. This line is produced through a forbidden transition emitted by low-density and warm gas (T$\sim10^{4}$K).
Thus, any disturbed kinematic, such as a broadening and asymmetry of the [OIII] line, is only due to the presence of strong bulk motions of ionized interstellar gas ( e.g. for local galaxies \citealt{Heckman+90,Veilleux+95,Lehnert+Heckman96,Soto+12,Westmoquette+12,Mullaney2013, Rodriguez-Zaurin+13,Bellocchi+13,Rupke+Veilleux13,Liu2013,Harrison2014,Zakamska2014,Cazzoli+14,Arribas+14,Cicone2016}, and at high redshift \citealt{Shapiro+09,Newman+12,Harrison+12,Cano-Diaz+12,Genzel+14,Forster-Schreiber+14,Brusa2015,Perna2015,Carniani+15}).
Also, [OIII]$\lambda5007$ emission is expected to lie in spectral regions free from strong stellar atmospheric absorptions.
In these respect, it is expected that the [OIII]$\lambda5007$ emission line can be used with great success to identify or confirm galactic winds.
We explore how the emission line profile changes as a function of key physical parameters: stellar mass, SFR and primary photoionization processes (SF, AGN). In doing, we analyse both the presence of a second broader Gaussian component and non-parametric variation of the line-profile.}

This work is organized as follows. In Section \ref{data}, we present our sample selection and physical properties. In Section \ref{method} we describe the method used to extract and analyze the [OIII]$\lambda 5007$ emission line from our stacked spectra. We present and discuss the main results in Section \ref{results} and finally we summarize our findings in Section \ref{conclusions}.
Throughout this paper, we assume the following cosmological parameters: $H_{0}=70$ km s$^{-1}$ Mpc$^{-1}$, $\Omega_{M}=0.3$ and $\Omega_{\Lambda}=0.7$

\section{Data sample}\label{data}


   \begin{figure}
   \centering
   \includegraphics[angle=-90, width=\hsize ]{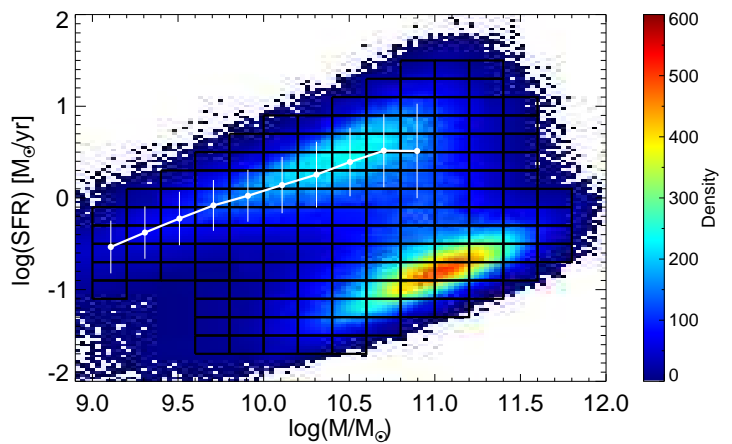}
   \caption{SFR-$M_{\star}$ plane for DR7 SDSS galaxies. The blacks boxes represent the fine grid used for the stacking of the total sample.The white line shows the position of the so-called "Main-Sequence" (MS) of star$-$forming galaxies. The MS is computed as the mode and the dispersion of the SFR distribution in stellar mass bins following the example of \citep{RenziniPeng2015}.}
              \label{FigPlane}%
   \end{figure}

   \begin{figure}
   \centering
   \includegraphics[angle=-90, width=\hsize ]{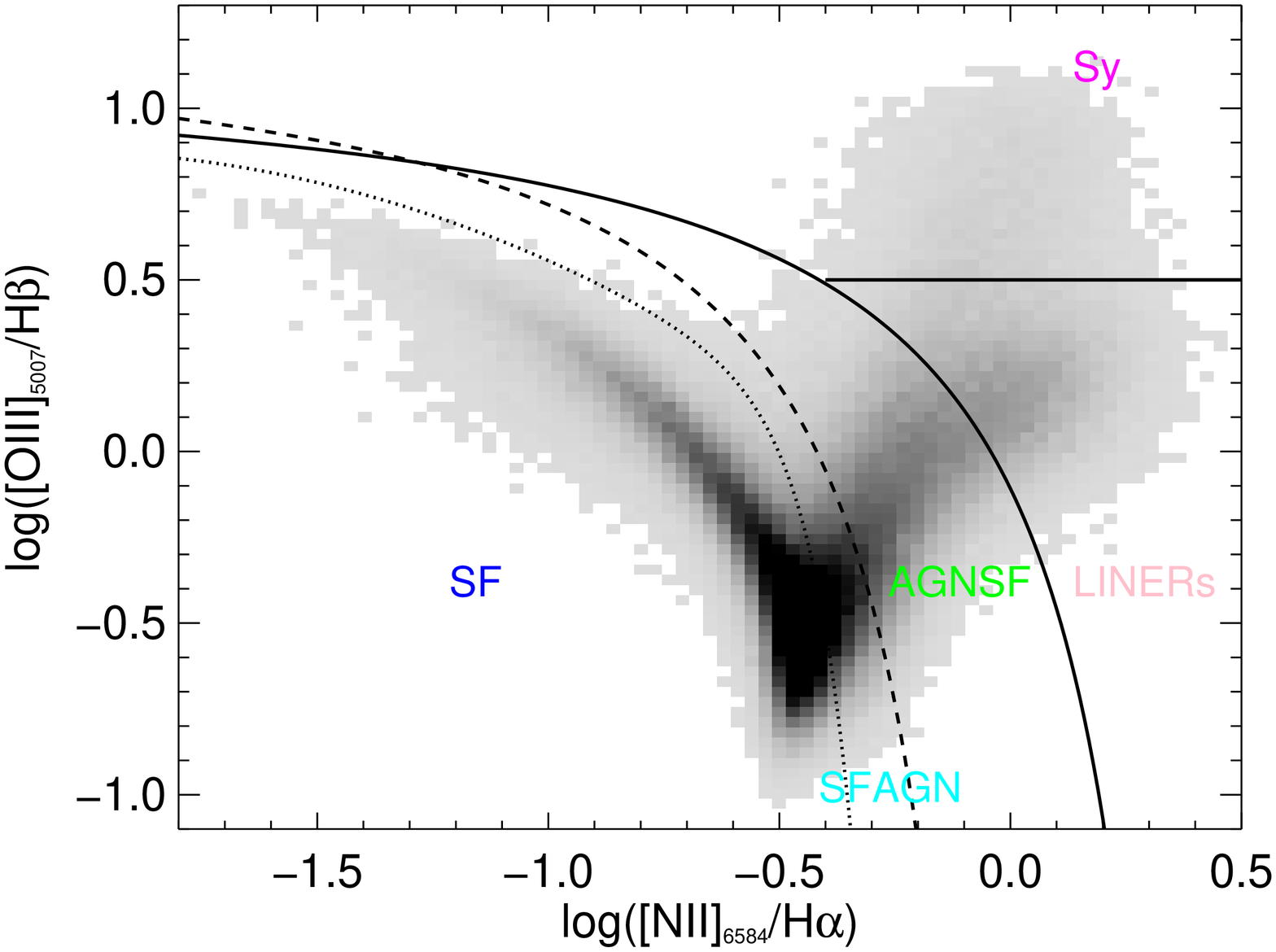}
   \caption{The distribution of the galaxies in our sample in the BPT line-ratio diagram.
The solid curve is the theoretical demarcation of \citet{Kewley2001} , that separates star-forming galaxies and composites from AGN. The dashed (\citealp{Kauffmann2003b} ) and dotted \citet{Stasinska2006}  curves indicate the empirical division between composite SF-AGN and AGN-SF and the pure star-forming galaxies, respectively (see text for more details). The horizontal line at $\log$ ([OIII]/H$\beta$) $= 0.5$ is the demarcation criteria between TYPE 2 (or Seyfert, Sy) and LINERs galaxies showed in \citet{Kewley2006}.}
              \label{FigBPT}
   \end{figure}

   \begin{figure}
   \centering
   \includegraphics[angle=-90, width=\hsize ]{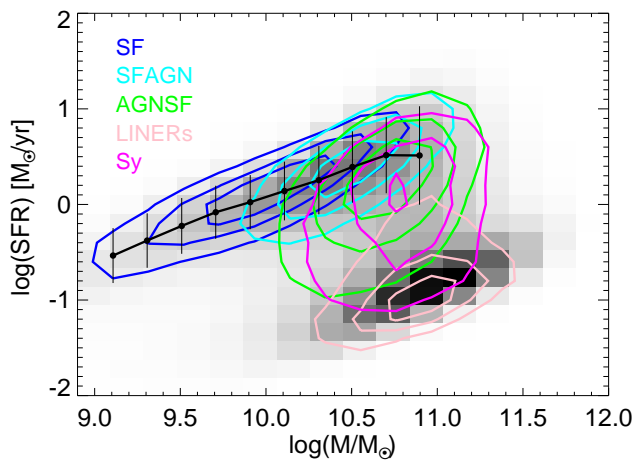}
   \caption{Location of SF, SF$-$AGN, AGN$-$SF, LINERs, TYPE 2 (or Sy) galaxies in the SFR-$M_{\star}$ plane. 
   From outside to inside the contours encompass $25$, $50$ and $75$ per cent of the data points. The black line shows the mode and dispersion of the MS.}
              \label{pianoBPT}
   \end{figure}

\subsection{SDSS spectra}
The galaxy sample analyzed in this paper is drawn from the Sloan Digital Sky Survey, (SDSS, \citealp{York2000}). In particular, we use the spectroscopic catalog containing $\sim 930,000$ spectra belonging to the seventh data release (DR7, \citealp{Abazajian2009}). We use only objects included in the Main Galaxy Sample (MGS, \citealp{Strauss2002}) which have Petrosian magnitude $r<17.77$ and redshift distribution extends from $0.005$ to $0.30$ , with a median z of $0.10$. The spectra cover a wavelength range  from $3800$ to $9200$ \AA. They are obtained with $3''$ diameter aperture fibers that, in the adopted cosmology, corresponds to $\approx 0.31-13.36$ kpc for the redshift range $z=0.005-0.3$. The instrumental resolution is $R \equiv\lambda / \delta \lambda \sim 1850-2200$ and a mean dispersion of $69$ km s$^{-1}$ pixel$^{-1}$. Further details concerning the DR7 spectra can be found at http://www.sdss.org/dr7/. 

\subsection{Star formation rates and stellar masses}
We adopt the star formation - stellar mass plane (hereafter SFR-$M_{\star}$) in order to classify galaxies in their fundamental properties. In order to define the position for each galaxy in the SFR-$M_{\star}$ plane, we use the SFR and $M_{\star}$ measurements taken from the MPA-JHU catalog\footnote{http://www.mpa-garching.mpg.de/SDSS/DR7/}. 
The stellar masses are obtained from a fit  to the spectral energy distribution (SED) by using the SDSS broad-band optical photometry (see \citealp{Kauffmann2003a} for more exhaustive details). The SFR measurements are based on the \cite{Brinchmann2004} approach. They use the H$\alpha$ emission line luminosity to determine the SFRs for the star forming galaxies. For all other galaxies, that have emission lines contaminated by AGN activity or not measurable emission lines, the SFRs are inferred by D4000-SFR relation \citep[e.g.][]{Kauffmann2003a}. All SFR measures are corrected for the fiber aperture following the approach proposed by \cite{Salim2007}. 
 
We apply a stellar mass cut at $\log(M/M_{\odot}) \geq 9.0$ to limit the incompleteness in the low mass regime (see also \citealp{Morselli2016}).
In this way, we ended up with a global sample of $\sim 600000$ galaxies.

{{The galaxy sample is shown in the SFR-stellar mass plane in Fig.\ref{FigPlane}. The color code is according to the number density of galaxies per bin of SFR and stellar mass. We overplot also the Main Sequence od star forming galaxies (MS hereafter), estimated as the peak (mode) of the distribution in the star forming galaxy region, similarly to \cite{RenziniPeng2015}. }}

\begin{table}
\small
\centering
\begin{tabular}{l|c|c|c}

\hline
\hline
     subsample   & number       & percent        & N Bins           \\
\hline
TOT           &$621990$     & $100\%$  &    $148$         \\
SF             &$128258$     & $20.6\%$  &    $88$    \\
SF-AGN    &$46081$       & $7.4\%$    &    $81$      \\
AGN-SF    &$69421$       & $11.2\%$   &    $119$          \\
LINERs     &$34640$       & $5.6\%$    &     $99$        \\
TYPE 2     &$10679$       & $1.7\%$    &     $77$        \\
unClass    &$332911$       & $53.5\%$  &     $132$          \\
\hline
\hline
\end{tabular}
\caption{Basic data about the subsamples discussed in the text.}
\label{tab:InfoSample}
\end{table}
\normalsize

\subsection{BPT classification}\label{bpt}
Emission line diagnostic diagrams are a powerful way to probe the nature of the dominant ionizing source in galaxies. \citeauthor*{BPT}(1981, BPT) and after them\cite{VeilleuxOsterbrock1987} demonstrate that it is possible to distinguish normal star-forming from AGN dominated galaxies by considering two pairs of emission lines ratios. 
The MPA-JHU catalog also includes, for each single spectrum, the flux measurements of [OIII]$\lambda 5007$, H$\beta$, H$\alpha$ and [NII]$\lambda 6584$ emission lines.
 As showed in \cite{Stasinska2006}, the galaxies that lie in the left side of the \cite{Kauffmann2003b} demarcation line includes also objects that have an AGN contamination. In order to better segregate the purely star-forming galaxies from AGN hosts, we refine the BPT classification of our sample instead to using the selection criteria performed by \cite{Brinchmann2004}.
By using the two optical line ratios:[OIII]$\lambda$5007/H$\beta$ and [NII]$\lambda$6584/H$\alpha$, then, we define new galaxies subsamples on the basis of the prevalence of different photoionization processes. All galaxies with no or very weak emission lines ($S/N<4$) are not classified in the BPT diagram and we call these objects "unClass".
For the lines with $S/N > 4$, we adopted the following classes of emission line nebulae: 
\begin{description}
\item[SF] Pure star-forming galaxies, objects with emission line ratio below the \cite{Stasinska2006} curve.
\item[SF$-$AGN] The objects whose emission lines are due primarily to star formation activity but that have also a second minor component due to AGN presence. They are located between the \cite{Stasinska2006} and \cite{Kauffmann2003b} demarkation lines. 
\item[AGN$-$SF] The composite transition region objects that lying inside the region defined by \cite{Kauffmann2003b} and \cite{Kewley2001} curves.
\item[Type 2 (or Seyfert galaxy, Sy) AGN and LINERs] All the objects located above the diagnostics outlined of \cite{Kewley2001} and separated in Seyfert galaxies and and low-ionization nuclear emission-line regions (LINERs) with the demarcation criteria showed in \cite{Kewley2006}, $\log([OIII]/H\beta) = 0.5$.
\end{description}

The corresponding diagnostic diagram is shown in Fig. \ref{FigBPT}.
We split the total sample in six classes:  SF, SF$-$AGN, AGN$-$SF, LINERs, TYPE 2 and unClass (cf. Table \ref{tab:InfoSample}). 
{In Fig. \ref{pianoBPT} we report the contour levels at 25\%, 50\% and 75\% of the distribution for the five BPT classes, labeled in different colors, in the SFR-$M_{\star}$ plane.}

{Throughout the paper we will also compare the above BPT classes with the TYPE1 AGN defined in \cite{Mullaney2013}. Due to the dominating AGN contribution, a measurement of the SFR and stellar mass derived from the spectra and optical broad band photometry is not available for such class of galaxies. Therefore, they can not be placed in the SFR-stellar mass plane.}

\begin{figure*}
\centering
\includegraphics[angle=-90, width=\hsize ]{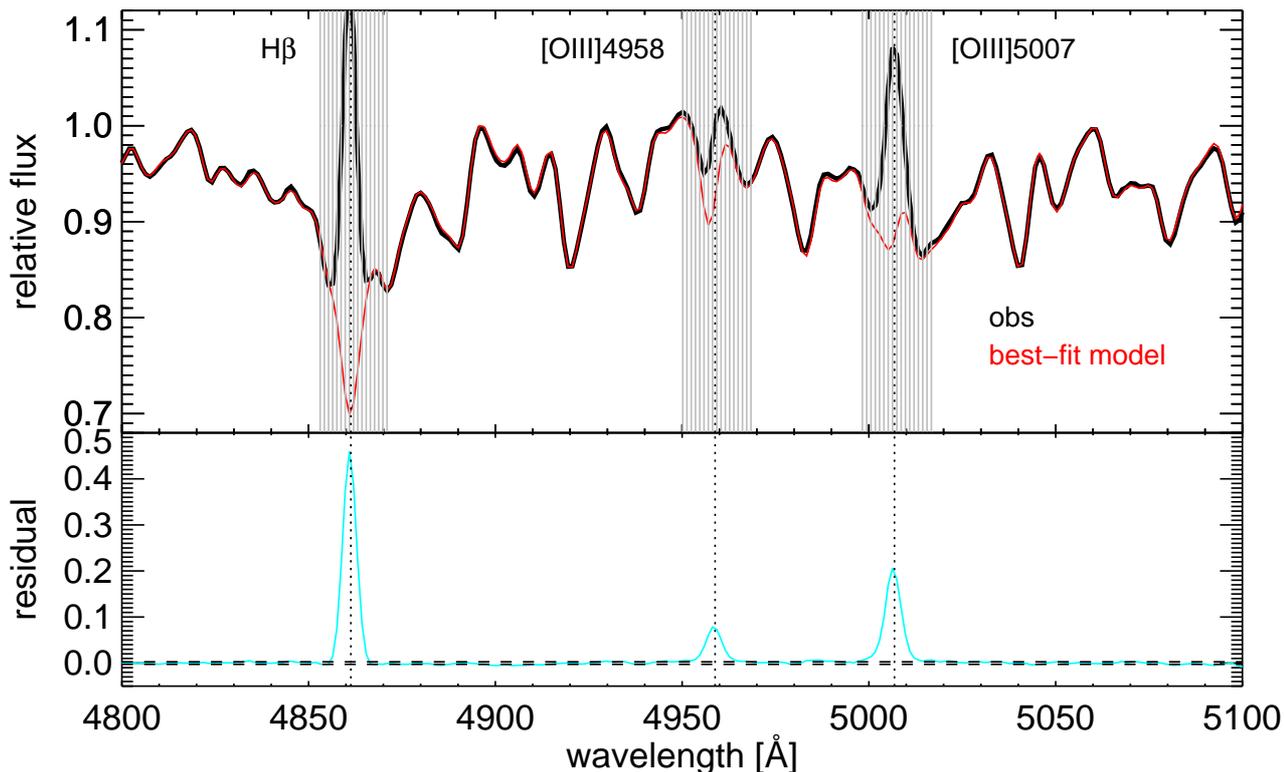}
\caption{Example of our continuum fit and subtraction performed for the stacked spectrum whit SFR=$10^{0.2} M_{\odot} yr^{-1}$ and $M_{\star}=10^{10.5} M_{\odot}$. The top panel shows the observed stacked spectrum (black line) and our best-fit stellar continuum model (red line). The light grey-shaded regions indicate the wavelength range where the H$\beta$, [OIII]$\lambda4959$ and [OIII]$\lambda5007$ emission lines are located. The bottom panel shows the residual spectrum (cyan line) and the level of fluctuations in the fit residuals (dashed line).}
\label{fig:esempio}
\end{figure*}   
\begin{figure*}
\centering
\includegraphics[angle=-90, width=\hsize ]{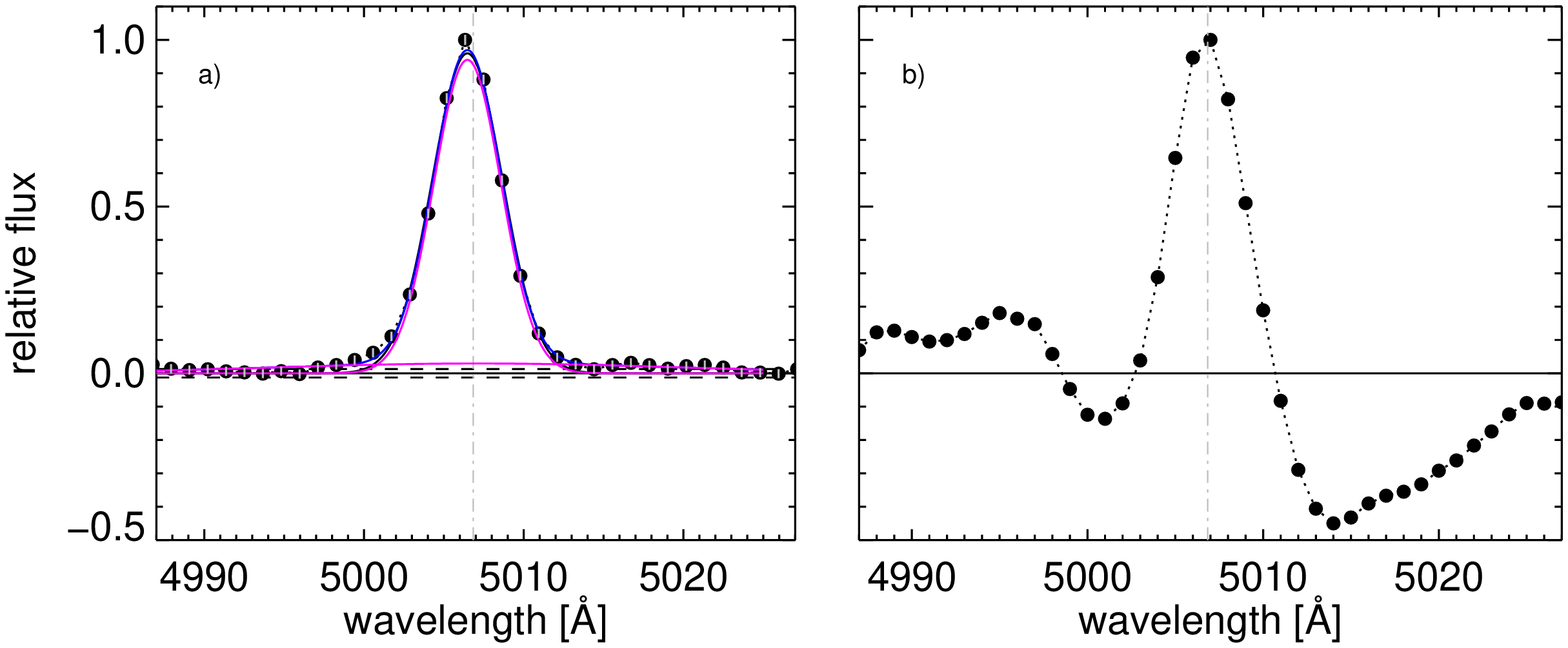}
\caption{Average [OIII]$\lambda 5007$ profile of the galaxy subsample whit SFR=$10^{0.2} M_{\odot} yr^{-1}$ and $M_{\star}=10^{10.5} M_{\odot}$. In panel a) we show the emission line derived with our method. In panel b) we show the emission line for the same SFR and $M_{\star}$ galaxies obtained by the method of \citep{Mullaney2013}. The black symbols are the observed flux. The magenta lines illustrates the two-Gaussian component and the blue curve shows the combined fit. The level of  scatter in the residuals of our fit is shown with the horizontal dashed lines. The vertical gray line mark the rest-frame position of the [OIII] line.}
\label{fig:confronto_Mullaney}
\end{figure*}   

\section{Method}\label{method}
In this section, we describe how we measure the properties of the [OIII]$\lambda 5007$ emission line. 

\subsection{Stacked spectra}
The auroral [OIII]$\lambda 5007$ emission line can be very faint and typically undetectable in most SDSS galaxy spectra. To reduce the contribution of random fluctuations in the measured flux and then improve the signal-to-noise ratio (SNR) we perform our analysis in stacked optical spectra. In particular, we use the median stacked spectra taken from Concas et al. in prep. In brief, we divide the SFR-$M_{\star}$ parameter space into small bins, shown in Fig.\ref{FigPlane}.
The boundaries of this grid together with the abundance of sources per bin, are chosen to provide a fine sampling of the SFR-$M_{\star}$ plane and at the same time to have good statistics in each bin. We adopt bins of  $\Delta \log(M/M_{\odot}) =0.2$ and $\Delta \log(SFR) =0.2$ dex for the total sample and larger bins for the analysis of the individual BPT classes, where the statistics is reduced.
We request a minimum of $50$ galaxies in each bin.
The galaxy spectra are first corrected for the foreground Galactic reddening using the extinction values from \cite{Schlegel1998} then they are transformed from vacuum wavelengths to air and shifted to the rest frame. We normalize each spectrum to the stellar continuum with the mean flux from $6400$ \AA to $6450$ \AA, where the spectrum is free of strong emission and absorption lines. Finally, the rest-frame spectra in each bin are stacked together to produce a single median spectrum. 
We obtain $148$ galaxy stacked spectra for the total sample and $88$, $81$, $119$, $99$, $77$  and $132$ for the subclasses described in section \ref{bpt}. 
The different numbers of stacked spectra between the total sample and the subsamples is due to the fact that towards the quiescence region less and less galaxies can be classified in the BPT diagram. In such region of the SFR-M$_{\star}$ plane we analyze the [OIII]$\lambda5007$ profile only in the stacked spectrum of the total sample.

{{The error of the stacked spectra is obtained through a bootstrapping analysis. The statistical error obtained in this way depends on the number of spectra used in the stacking analysis in each bin. However, we check the the error is very stable from the most populated bins ($\sim 4000$ galaxy spectra) to the less populated ones ($50$ galaxies).}}


\begin{figure}
\centering
\includegraphics[angle=-90, width=\hsize]{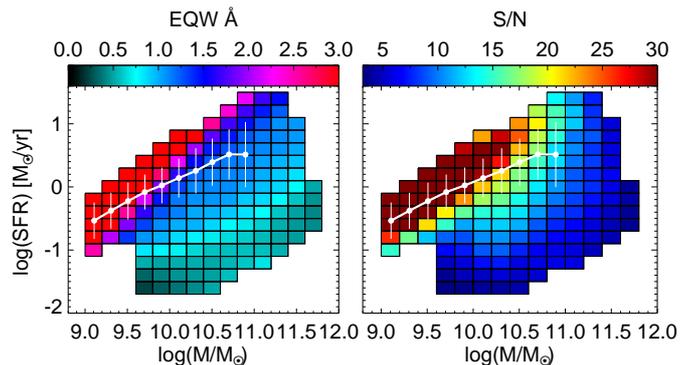}
\caption{EQW (left panel) and total line signal-to-noise ratio, SNR (right panel) in the SFR-M$_{\star}$ diagram for the total sample. The white line shows the mode and dispersion of the MS. The galaxy bins whit total emission line SNR $< 8$ are plotted in grey color.}
\label{fig:eqw_SNR}
\end{figure}   

\subsection{Fitting the stellar continuum}
In order to measure reliably also the weak emission lines emitted by the ionized gas, the stellar continuum must be properly removed. To this purpose, we use the penalized pixel-fitting (pPXF) algorithm, which is a publicly available IDL code, developed by \citet{Cappelalri_Emsellem2004} to find the best fit stellar continuum and separate the nebular emission lines in each stacked spectrum.
In brief, pPXF is able to parameterized the line-of-sight velocity distribution (LOSVD) through a Gauss-Hermite expansion of the absorption-line profile by fitting the stellar continuum with a set of linear combination of simple stellar population (SSP) input model spectra.
In the pPXF analysis we adopt a library of template spectra based on the stellar population models from \citet{BC03}, hereafter BC03. BC03 models are available at a resolution of $3$ \AA\ FWHM in the wavelength range between $3200-9500$~\AA, which is very similar to the one of SDSS spectra ($\approx1800-2000$ between $3800$ to $9200$~\AA). Our templates include simple stellar population with age $0.01 \leq t \leq 14$ Gyr and four different metallicity, $Z/Z_{\odot}=0.2, 0.4, 1, 2.5$ by assuming a \cite{Chabrier2003} initial mass function (IMF).
We perform the pPXF analysis for each stacked spectrum in the wavelength range : [$4800, 5050$] \AA, where the  H$\beta$, [OIII]$\lambda 4959, 5007$ emission lines are located. The results are a best-fit stellar continuum. For each galaxy bin we subtract the best-fit stellar continuum from the observed stacked spectrum. This "residual" spectrum is used for any analysis of [OIII]$\lambda$5007 emission line features. {{ {An example of the fitting procedure result is shown in Fig. \ref{fig:esempio}, which shows the very good agreement of the observed and the model continuum over a large wavelength range in the [OIII] emission line region. In order to check the stability of the fitting procedure and to estimate the error of the residual spectrum, we apply a bootstrapping technique by performing the fit on the sample of bootstrapped stacked spectra in each SFR-stellar mass bin (see previous paragraph). The stability of the procedure is confirmed by the fact that the error of the residual spectrum is consistent or only slightly larger (at maximum 30\%) with respect to the error of the stacked spectra. As an example, the panel a) of Fig. \ref{fig:confronto_Mullaney} shows the residual spectrum in the [OIII]$\lambda 5007$ emission line region. The quality of our continuum fit is guaranteed by the low level of fluctuations in the fit residuals (dashed lines in Fig. \ref{fig:confronto_Mullaney}). The error of the residual spectrum is, then, used to estimate the SNR of the emission line in the residual stacked spectra.}

{For comparison we show also the result of the continuum subtraction method applied by \cite{Mullaney2013} on the TYPE1 AGN sample ( Fig. \ref{fig:confronto_Mullaney}, panel b). They use, in particular, the single continuum subtracted spectra , provided by the SDSS pipeline. While this method turns out to be reasonable for AGN spectra where the emission line has a very high SNR with respect to the continuum, it is not applicable to galaxy spectra with lower SNR emission lines, as in the considered case. }

\subsection{Measuring [OIII]$\lambda 5007$ emission line profiles}
We analyze the oxygen line shape with two different approaches: a) by fitting the line with a single and a double Gaussian to identify a possible second broader component with respect to the systemic one, and b) by adopting a non-parametric analysis. 
The two methods are complementary. The first one allows us to study separately the various components that determine the observed line, while the second procedure is independent to the particular fitting function and it is relatively insensitive to the quality of the data (see \citealp{Perna2015}, \citealp{Zakamska2014}, \citealp{Liu2013}).

\subsubsection{Profile fitting}\label{fit_method}
{We fit the [OIII] line profile in the residual spectrum with one and two Gaussian components by using an IDL MPFIT fitting code. In both cases we fit the line center, width and amplitude. }

{The single Gaussian fit allows to estimate the global line width, $\sigma_{[OIII]}$ and the SNR of the line. The observed $\sigma_{obs}$ of the line is the convolution of the real width of the emitted line and the instrumental resolution. To remove the instrumental effects we correct the $\sigma_{obs}$ whit $\sigma_{[OIII]}=\sqrt{\sigma_{obs}^{2} - \sigma_{inst}^2}$, where $\sigma_{inst}^{2}$ is the instrumental dispersion. For SDSS data the $\sigma_{inst}$ change as a function of wavelength, and it varies with the location of the object on the plate and the temperature on the night of the observations. Therefore, in order to use the correct $\sigma_{inst}$ for all the our stacked spectra, we use the instrumental resolution measured for each single spectrum from the ARC lamps provide by the MPA-JHU group. The mean $\sigma_{inst}$ in the [OIII] wavelength range for our sample is $\sim 60$ km s$^{-1}$.}

{The double Gaussian fit allows to estimate the significance of a second component and its line profile. We take the double Gaussian profile as the best fit for the [OIII] line profile when it leads to a reduction of the $\chi^{2}$ value by more than $30\%$. This is to avoid a misidentification of a second component when the double Gaussian fit provides two Gaussian components with consistent width and center and different amplitude. Indeed, in this case the sum of the two Gaussian would lead anyhow to a single Gaussian component. We check that a reduction of the $\chi^{2}$ value by at least 30\% is a good threshold to distinguish the need of a double Gaussian fit.}

{When the double Gaussian fit is retained as the best fit, we use it to estimate a) the SNR of the second component to check its significance, b) to compare the flux percentage of the second component with respect to the systemic contribution, c) to estimate the velocity shift of the centroid with respect to the systemic redshift and d) to estimate the line width of second and systemic component.}

{The errors on all measured quantities are obtained with a bootstrapping technique. The fit is repeated on the bootstrapping sample (see previous paragraph) to obtained the distribution of all measured quantities and so the dispersion as an estimate of the error.}

\begin{figure}
\centering
\includegraphics[angle=-90, width=\hsize]{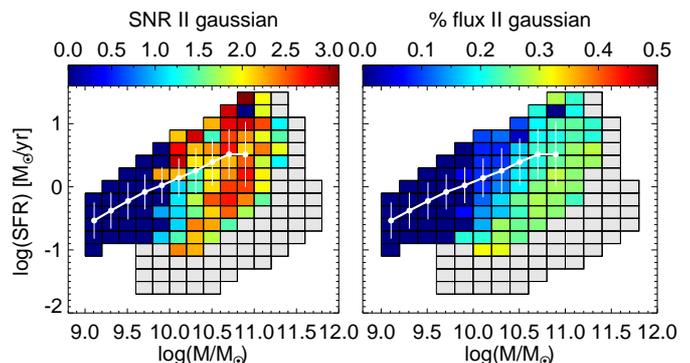}
\caption{Signal-to-noise ratio, (SNR) and flux enclosed in the second broader Gaussian component (right and left panel) in the SFR-M$_{\star}$ diagram for the total sample. The white line shows the mode and dispersion of the MS. The galaxy bins whit total emission line SNR $< 8$ are plotted in grey color.}
\label{fig:fluxII}
\end{figure}   

 %

\begin{figure}
\centering
\includegraphics[angle=-90, width=\hsize]{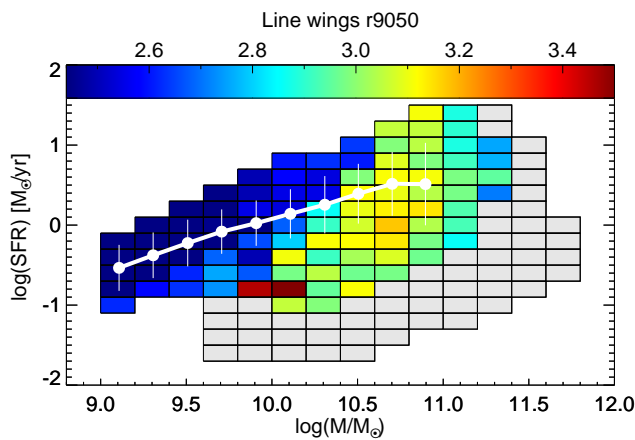}
\caption{Prominence of the line wings $r9050$ in the SFR-M$_{\star}$ diagram for the total sample. The white line shows the mode and dispersion of the MS. The bins below M$_{\star} = 10^{10.5}$ M$\sun$ have $r9050$ values consistent with 2.44 within $3 \sigma$. The galaxy bins whit total emission line SNR $< 8$ are plotted in grey color.}
\label{fig:r9050}
\end{figure}   

%
\begin{figure}
\centering
\includegraphics[angle=-90, width=\hsize]{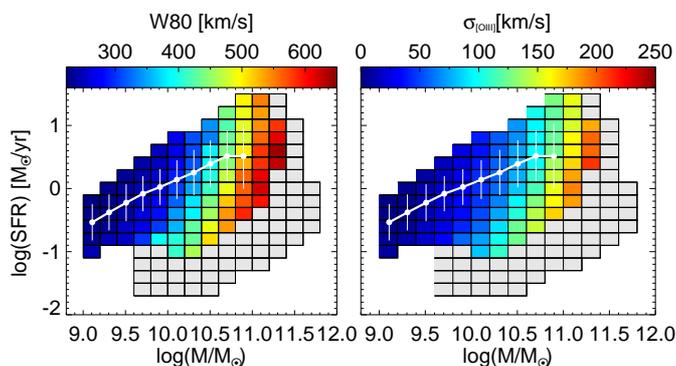}
\caption{Non-parametric $w80$ and $\sigma_{[OIII]}$ (left and right panels, respectively), in the SFR-M$_{\star}$ diagram for the total sample. The white line shows the mode and dispersion of the MS.}
\label{fig:line_width}
\end{figure}   

\begin{figure*}
\centering
\includegraphics[angle=-90 , width=0.33\textwidth]{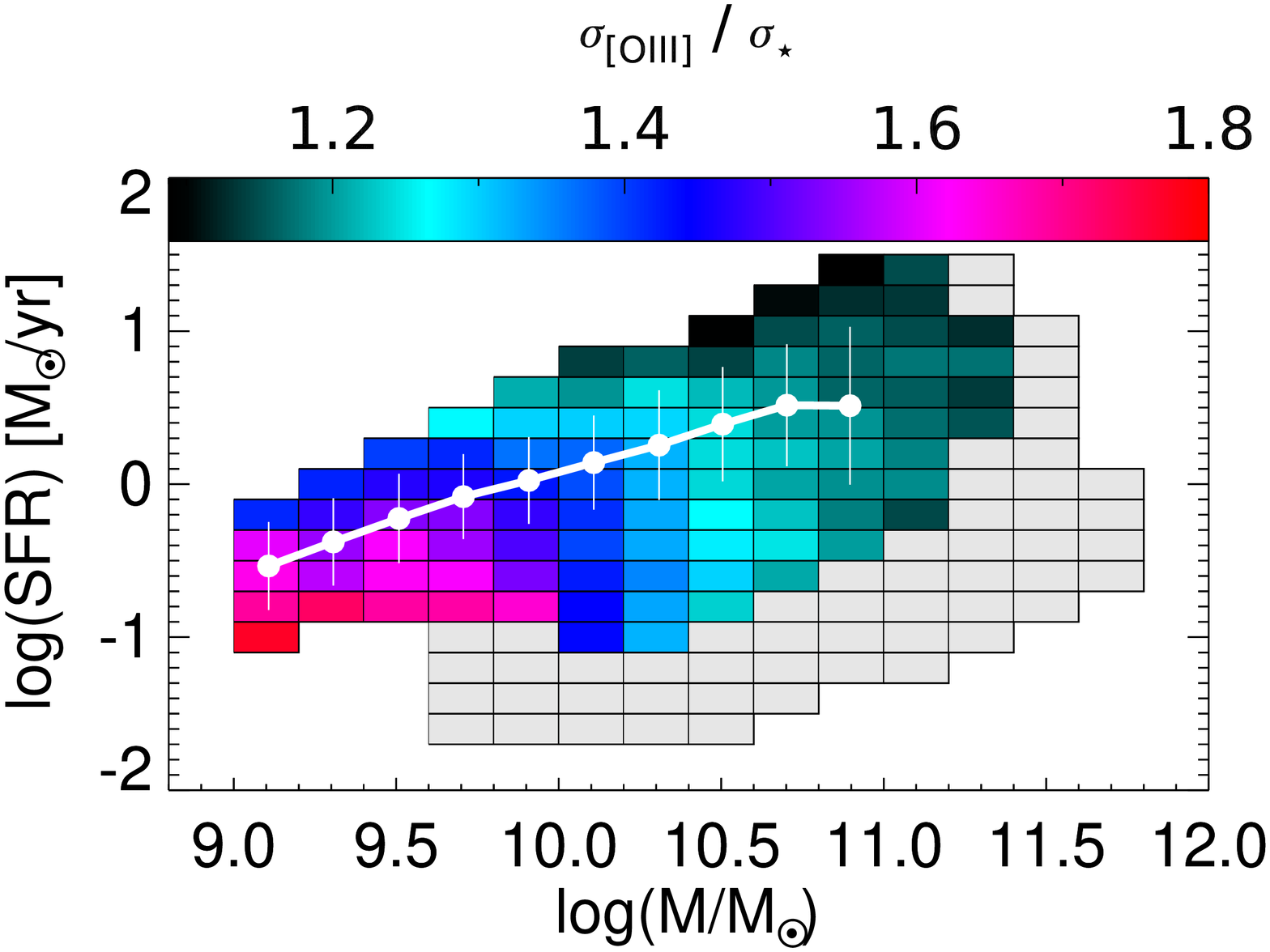}
\includegraphics[angle=-90 , width=0.33\textwidth]{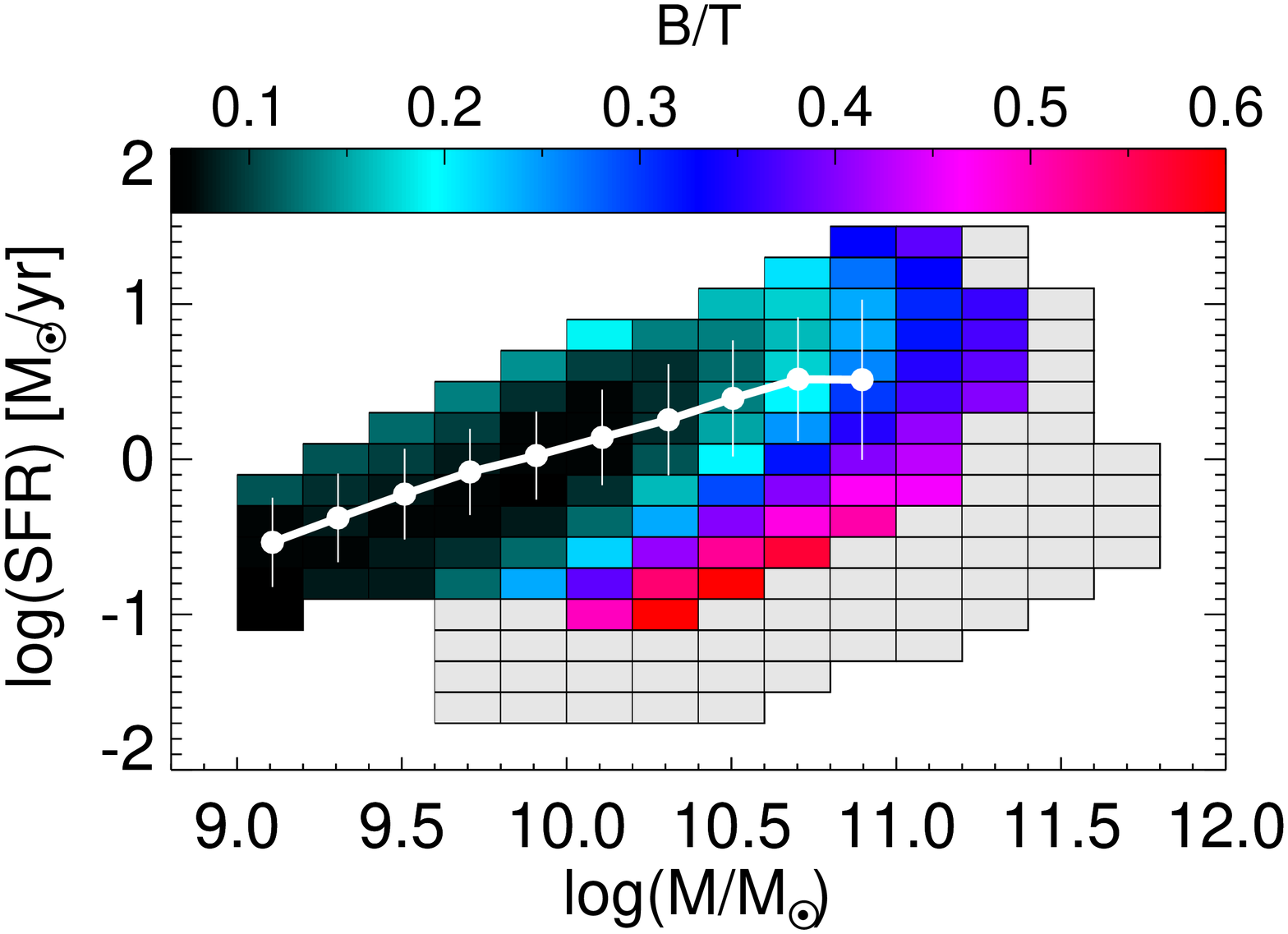}
\includegraphics[angle=-90 , width=0.33\textwidth]{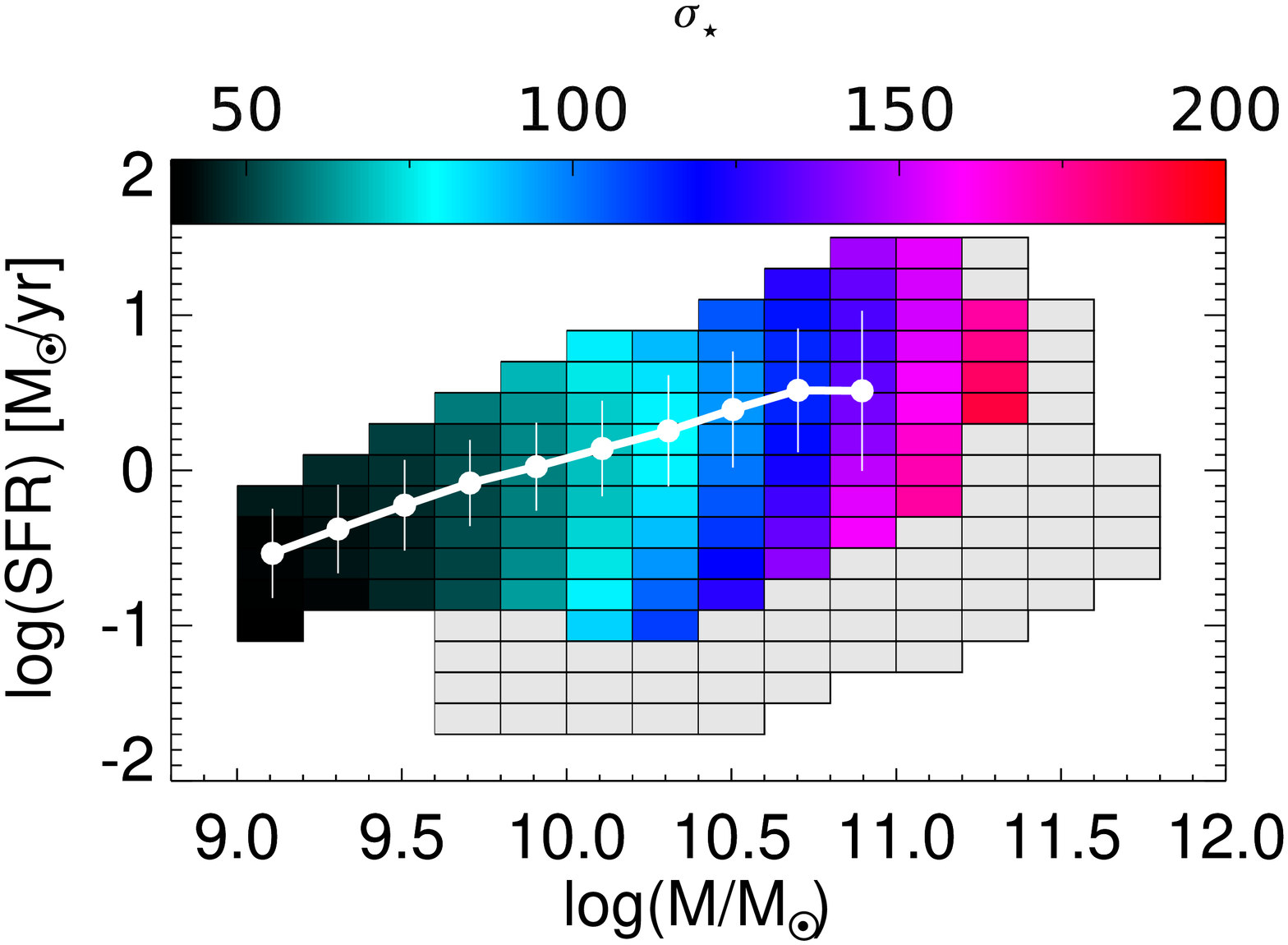}

\caption{{\it{Left panel}}: $\sigma_{[OIII]} / \sigma_{\star}$ in the SFR-M$\star$ plane. At high stellar mass the gas kinematics follows the velocity dispersion of the stellar component. {\it{Central panel}} B/T median values in the SFR-M$\star$ diagram. The bulge-disk decomposition is taken from \cite{Simard+11} catalogue. We use the values calculated with the r filter. {\it{Right panel}}: $\sigma_{\star}$ distribution in the SFR-M$\star$ plane. At low stellar masses the $\sigma_{\star}$ is below to the instrumental resolution. In all panels, the white line shows the mode and dispersion of the MS.}
\label{fig:planeBsuTr}
\end{figure*}   

\subsubsection{Non-parametric analysis}
To have a model independent measurement of emission line profiles we also apply a non-parametric approach.
This approach is commonly used in AGN outflows studies \citep{Liu2013,Rupke+Veilleux13,Zakamska2014,Harrison2014,Brusa2015,Perna2015}. 
{Briefly, we construct the cumulative flux of the line as a function of velocity: $F(v) =\int_{-\infty}^v \! f(v') \, \mathrm{d}v'$, in the observed spectrum without using any particular fitting function.}
Then, we describe the velocity width, asymmetry and the wings prominence of the [OIII] line by using the following non-parametric quantities: 
\begin{enumerate}
\item {Velocity width}.The  velocity width,  {$W80$} , is the velocity range that encloses $80$\% of the total flux. It is defined by $W80=v90-v10$, where $v90$ and $v10$ are the velocities at which $90$\% and  $10$\% of the line flux accumulates, respectively. For a purely Gaussian velocity profile the $W80$ is proportional to the standard deviation ($\sigma$) and  full width at half maximum (FWHM), as shown in the following equation, $W80 = 2.563 \times \sigma = 1.088 \times $FWHM. Values of $W80$ are given in km s$^{-1}$.
\item {Asymmetry}. The dimensionless parameter  {$R$} $=((v95-v50)-(v50-v05) ) / (v95-v05)$ gives a measure of the asymmetry of the velocity profile relative to the median velocity. In a perfect symmetric profile R is $R=0$
\item {Line wings}. The prominence of the line wings in the profile is the non-parametric analog of the kurtosis, with  {$r9050$}$= W90/W50$, where $W90$ and $W50$ are the width comprising $90$\% and $50$\% of the flux, $W90=v95-v05$ and $W50=v75-v25$. In a Gaussian profile $r9050$ is equal to $2.4389$, the $r9050$ increases in profiles with more extended wings.
\end{enumerate}

{The error on each quantities is estimated, as in the previous case, via bootstrapping analysis. Each quantity is measured in the bootstrapping sample related to each residual spectrum in order to estimate the dispersion of the distribution as a measure of the error. This is done, in particular, to check if each quantity deviates more then 3$\sigma$ from the value corresponding to a Gaussian distribution with the same width of the observed line. For this purpose, we use the measure of the global line width estimated with the single Gaussian line profile, as explained in the previous paragraph.}

\section{Results}\label{results}
In this section we show our results for the total sample and for the five BPT classes: SF, SF$-$AGN, AGN$-$SF, LINERs and TYPE2 AGNs. We show also the comparison with the TYPE1 AGN sample of \cite{Mullaney2013}.

\subsection{[OIII] line in the global sample} 
{We analyse the line flux and shape of the total sample as a function of position in the SFR-M$\star$ diagram.
Fig. \ref{fig:eqw_SNR} show the distribution of the [OIII]$\lambda5007$ equivalent width (EQW, left panel) and corresponding signal-to-noise ratio (right panel), SNR, of the total emission line in the SFR-M$\star$ plane. As expected, the MS region is populated by the higher EQW values and higher SNR, while
towards the passive region, the line is intrinsically weak, with EQW $\leq 1$ \AA\ and low SNR, SNR$\leq 8$. In order to ensure a robust  and accurate measurement of the total emission line shape, we impose a SNR limit on the [OIII] line of 8. In later figures of the paper, all bins whit total emission line SNR < 8 are plotted in gray color and they are not considered in the analysis.}

{We use the results of the best Gaussian fit, single or double, to check the significance of the second Gaussian component. When a single Gaussian turns out to be the best fit, we set the SNR of the second component equal to zero. When the best fit is provided by a double Gaussian, we estimate the SNR of the second component and the percentage of line flux encapsulated in that component. This is done in each bin of the SFR-stellar mass plane. The left panel of Fig. \ref{fig:eqw_SNR} show the SNR of the second Gaussian component, while the right panel shows the  percentage of flux encapsulated in it. 
Despite the very high SNR of the [OIII] line, the second broad Gaussian component is only marginally detected with a SNR$\sim$2.5-3 at masses above $\sim 10^{10} M_{\odot}$ and in a large range of SFR. Al lower masses, the [0III] line is perfectly consistent with a single Gaussian and no additional component is needed to fit the line profile. The flux encapsulated in the second broad component, though, is less than 10\% in most of the MS region, with the exclusions of the highest mass bins, where it reaches a value of 20\%. It reaches a similar percentage (20-25\%) also in the region below the MS, the so called green valley. }


{This is also confirmed by the non-parametric analysis.  The vaules of the line wings parameter, $r9050$, as a function of the position in the SFR-M$\star$ plane, is shown in Fig. \ref{fig:r9050}. The $r9050$ parameter is consistent with the value of a Gaussian function ($r9050\sim 2.44$) along and around the MS relation, up to stellar masses of $10^{10.5} M_{\odot}$. The kurtosis is exceeding, with poor significance ($\sim 2\sigma$), the Gaussian value up to values of $2.9-3$ in the mass range $10^{10.5-11}$ $M_{\odot}$, where the percentage of flux encapsulated in the broader component is of 20-25\%. For comparison, the kurtosis of a Lorentzian profile, with strong wings, is $6.31$.}
{In all bins, the [OIII] line appears to be symmetrical, with typical $R$ values always consistent with $0$ within 3$\sigma$ of significance.} 

Fig. \ref{fig:line_width} shows the distribution of the line width in the SFR-M$_{\star}$ plane, estimated either with the dispersion $\sigma_{[OIII]}$ or the analogous non-parametric $W80$ (right and left side, respectively). We observe a progressive increase of the line width with the galaxy stellar mass M$_{\star}$ at any SFR. This is expected since the [OIII] emission traces the galaxy potential well. We compare the value of $\sigma_{[OIII]}$ in any bin of the plane with the mean galaxy velocity dispersion, estimated from the absorption features due to the stellar component provided by the MPA-JHU public catalog. The left panel of Fig. \ref{fig:planeBsuTr} shows that at high stellar masses, the ratio between $\sigma_{[OIII]}$ and the galaxy velocity dispersion is consistent with $1$. Only at lower masses ($< 10^{10.5}$ $M_{\odot}$) the ratio increases to higher values. However in this region none of the previous indicators (flux enclosed in the second broader component, asymmetry $R$ or the kurtosis $r9050$) shows signature of a non Gaussian line profile. Thus, we ascribe such increase to two factors: a) the galaxies in this region tend to be pure disks, as shown by the mean B/T of $\sim 0.1$ as derived from Simard et al. (2011, central panel of Fig. \ref{fig:planeBsuTr}), thus the gas could show a different kinematics than the stellar component; b) in this region of the diagram, the MPA-JHU public catalog provide values of the stellar velocity dispersion lower than the SDSS resolution of $70 km/s$, which we assume as a lower limit. (right panel of Fig. \ref{fig:planeBsuTr}). 



\begin{figure*}
{\centering%
\begin{tabular}{@{}lll@{}}
\includegraphics[angle=-90,width=0.33\textwidth]{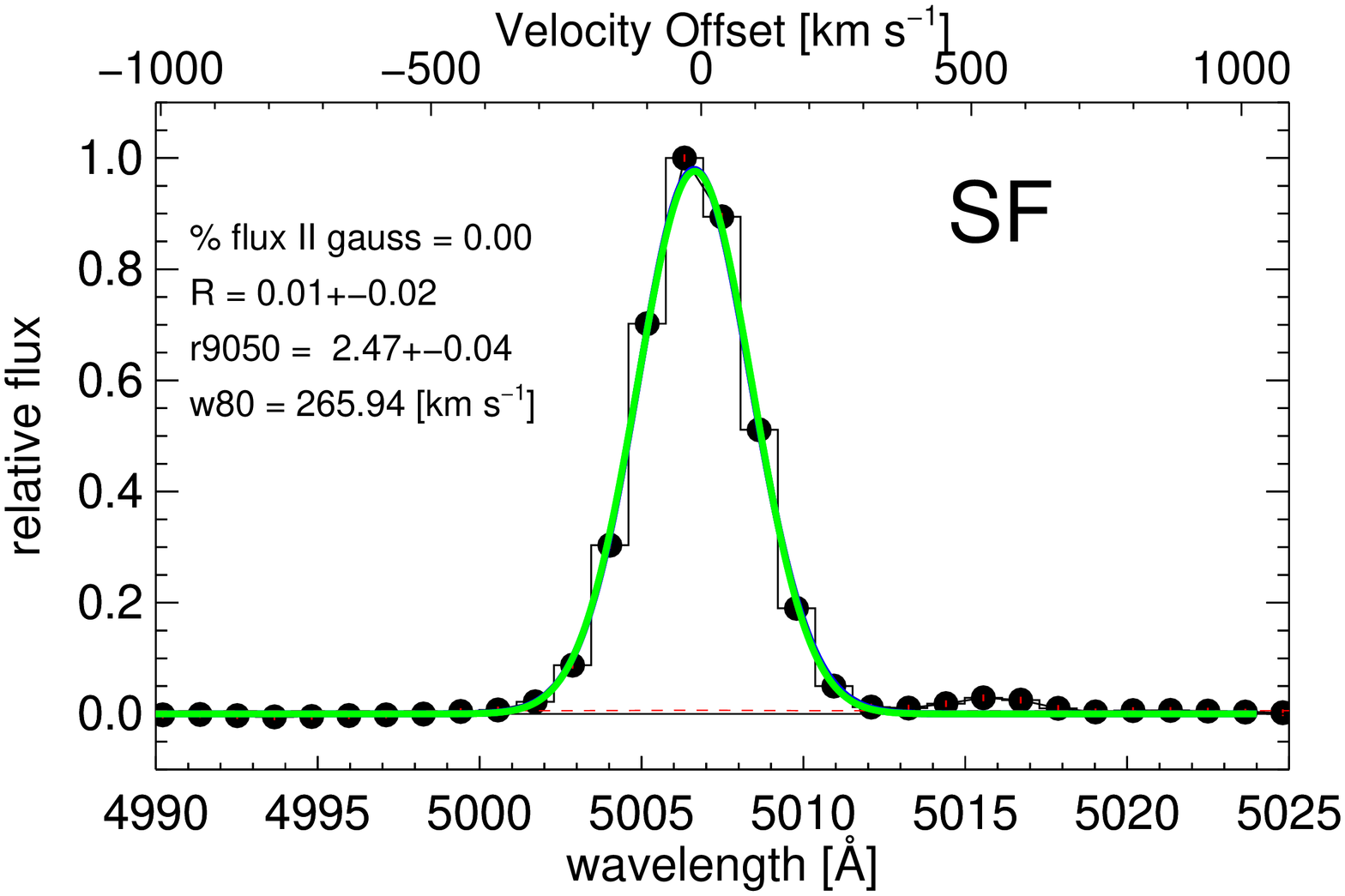}&
\includegraphics[angle=-90,width=0.33\textwidth]{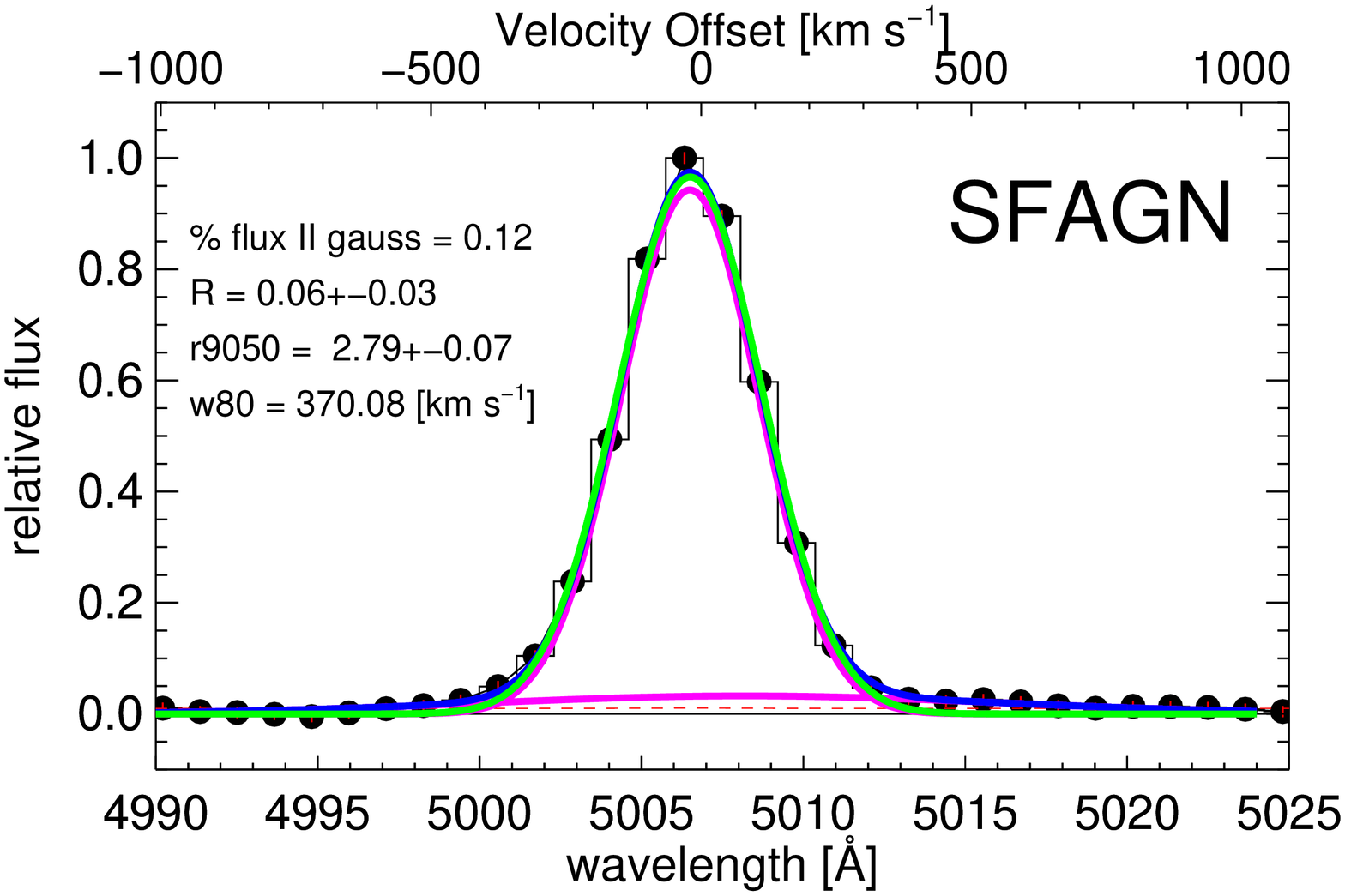}&
\includegraphics[angle=-90,width=0.33\textwidth]{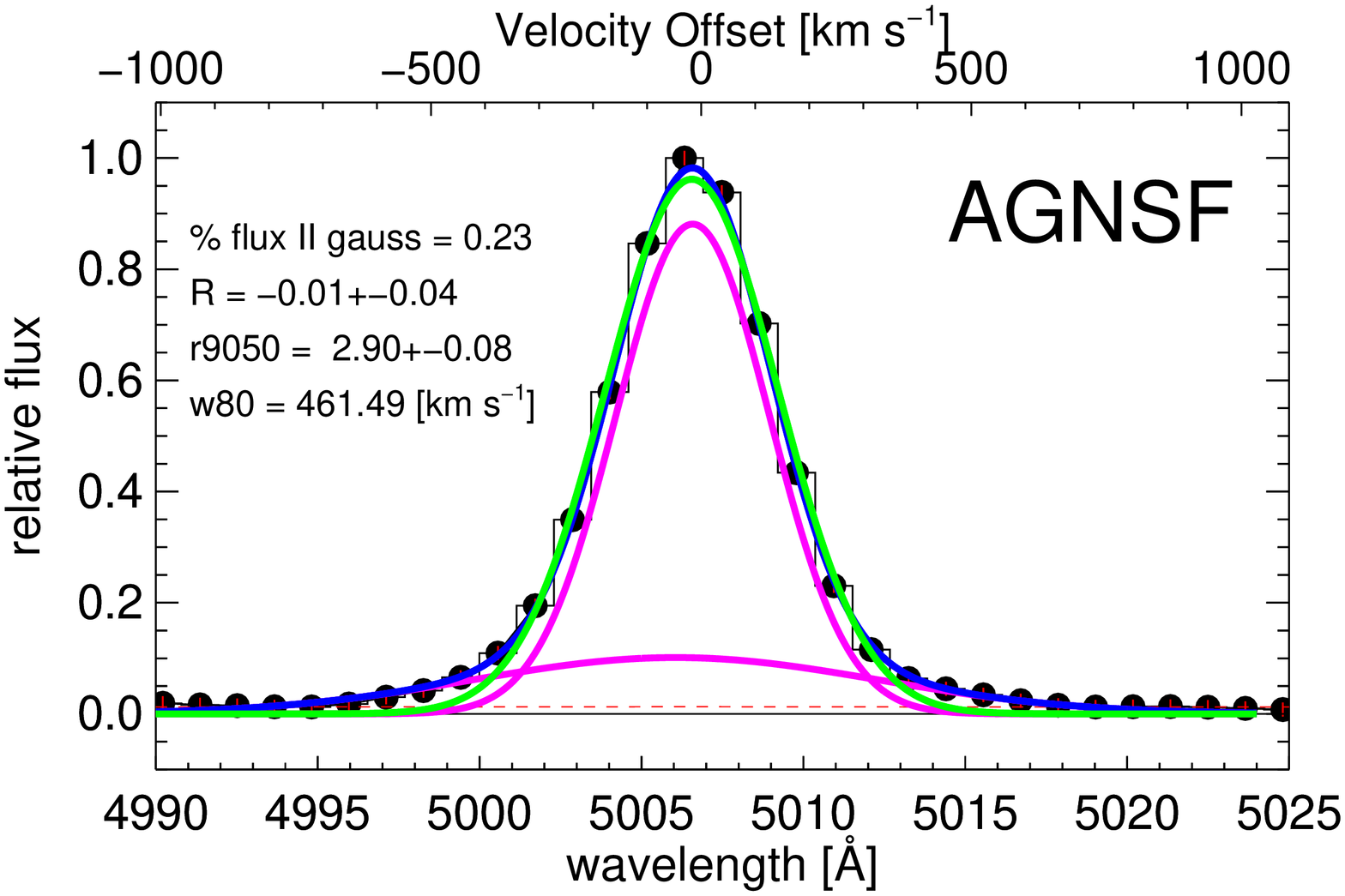}
\end{tabular}\par}

{\centering%
\begin{tabular}{@{}lll@{}}
\includegraphics[angle=-90,width=0.33\textwidth]{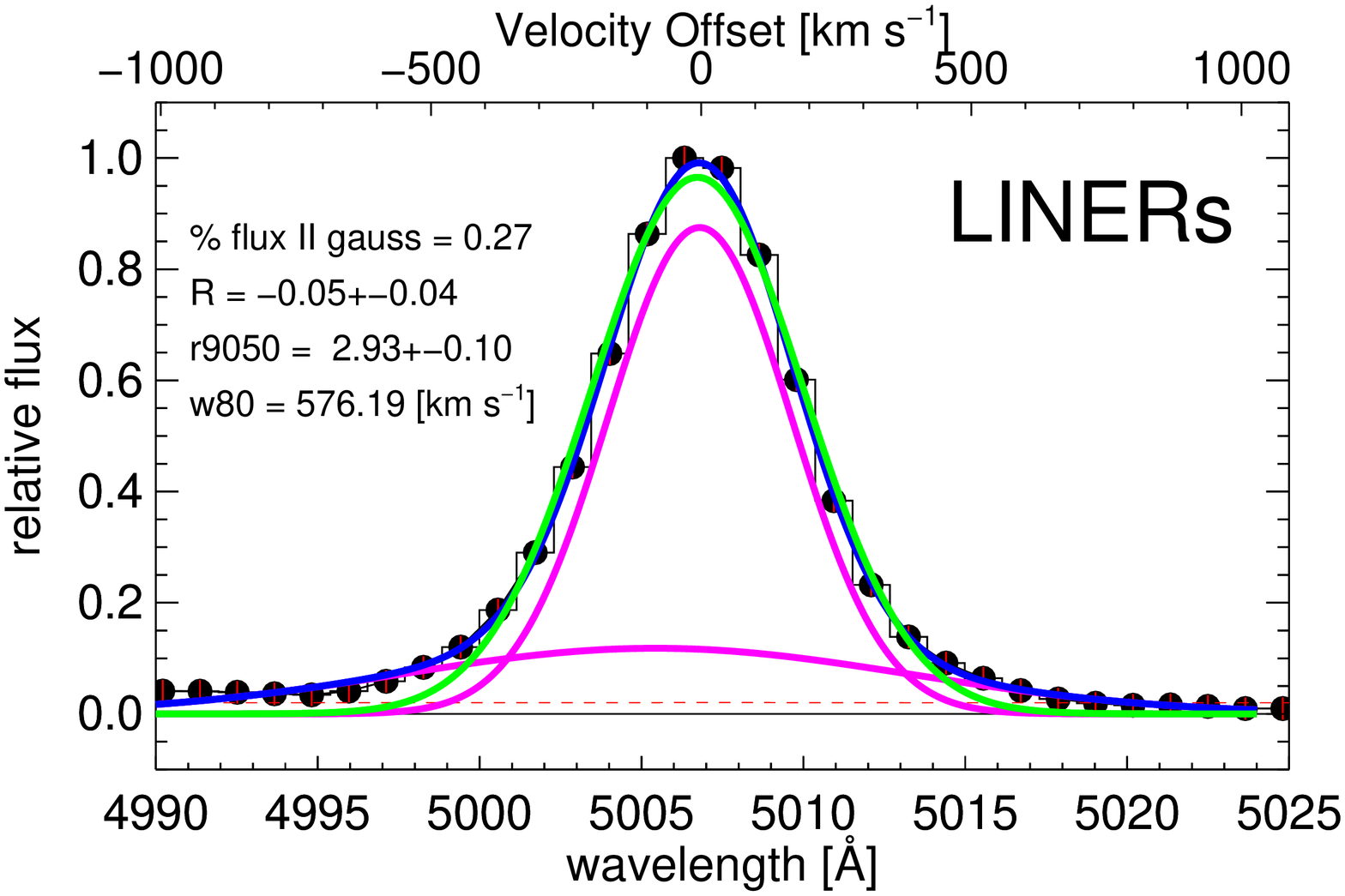}&
\includegraphics[angle=-90, width=0.33\textwidth]{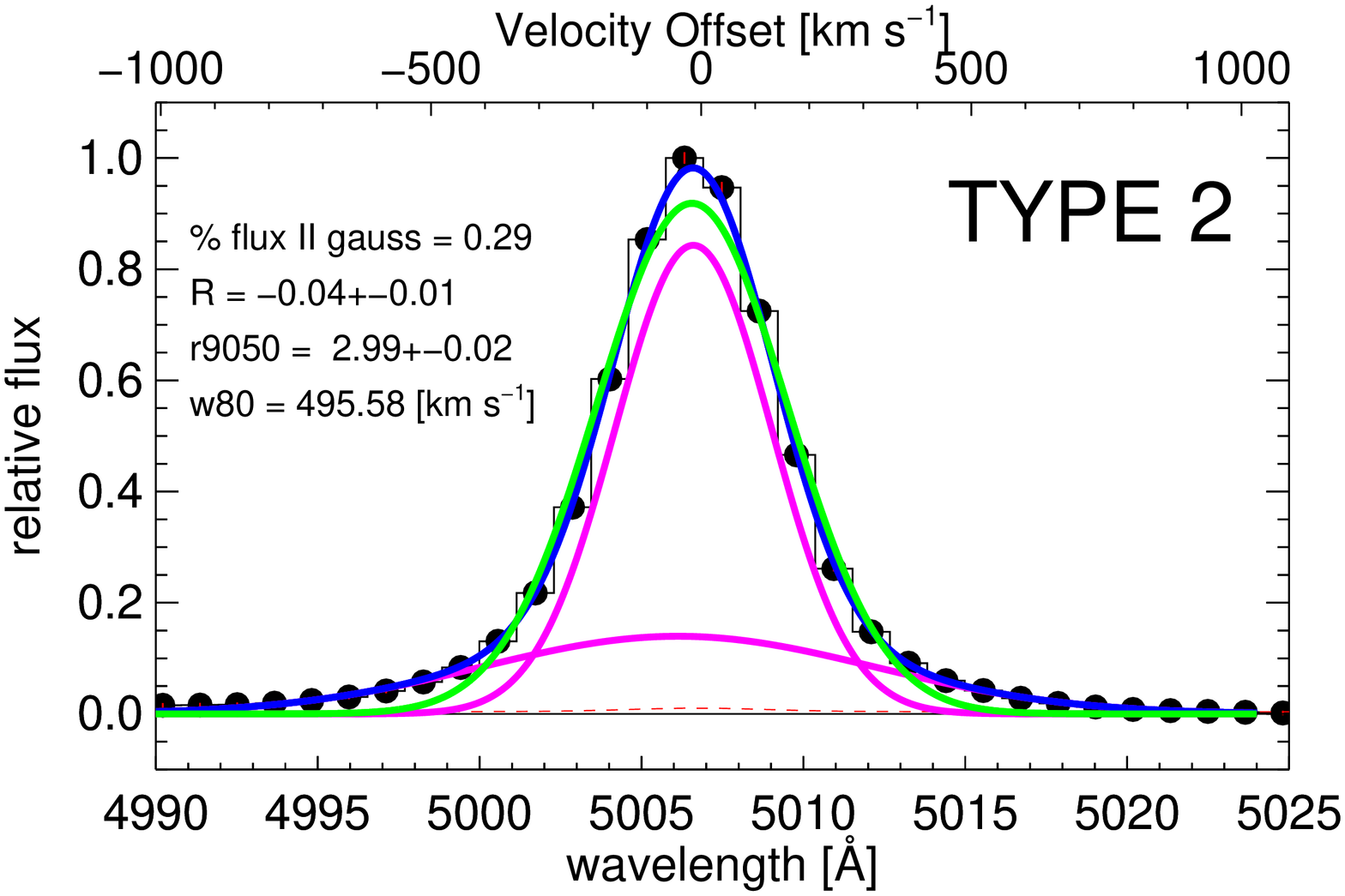}&
\includegraphics[angle=-90, width=0.33\textwidth]{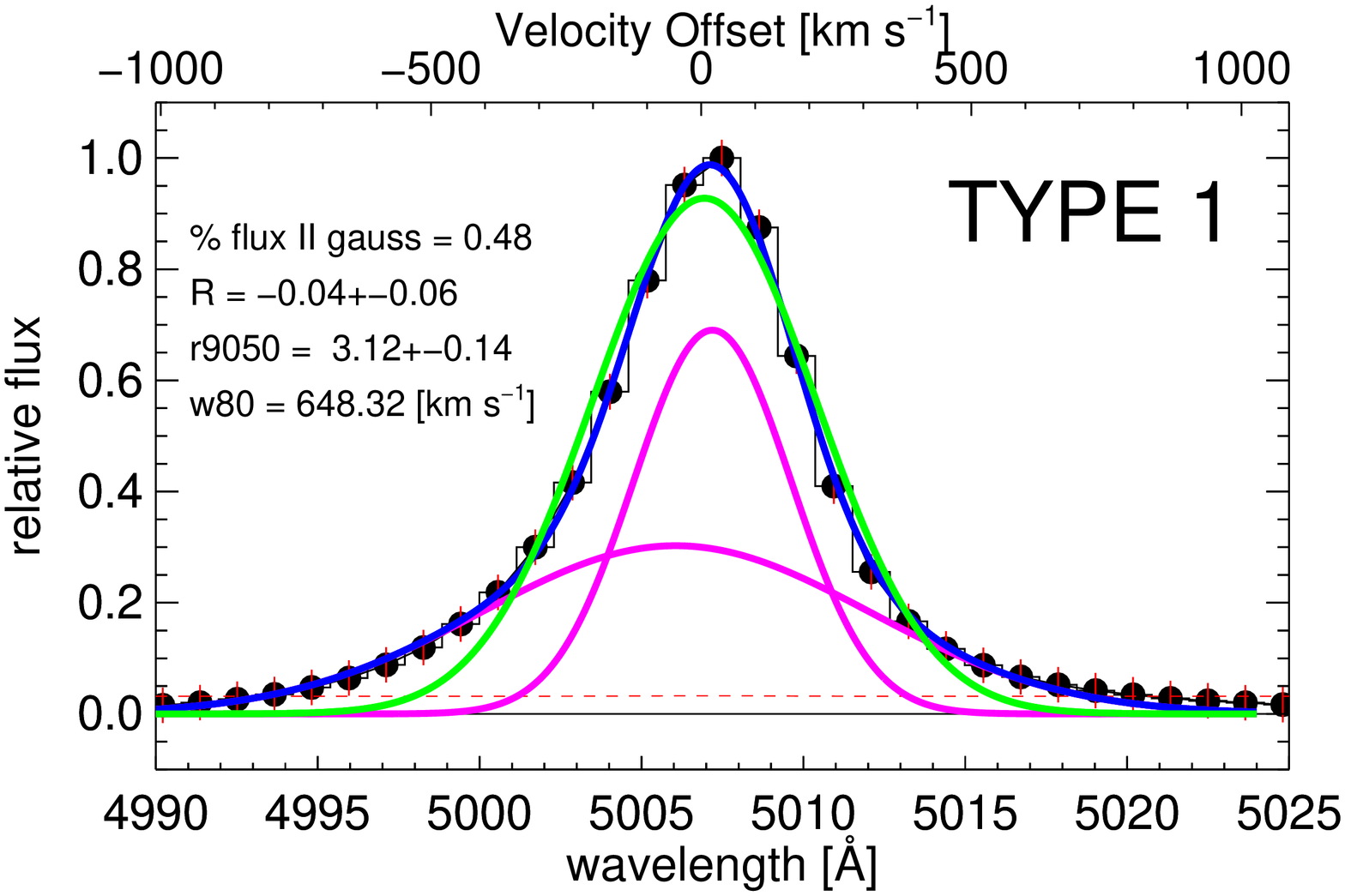}
\end{tabular}
\caption{Emission-line profile fits to composite spectra in different classes of photoionization processes: SF, SF$-$AGN, AGN$-$SF, LINERs, TYPE 2 and TYPE 1. The black symbols are the observed flux. The flux errors, in each point, are showed in red . The green line shows the single-Gaussian fit. The magenta lines illustrates the two-Gaussian component and the blue curve shows the combined fit. The level of scatter in the residuals of our fit is shown with the horizontal dashed red lines. The significance of a second broad and blue-shifted component (magenta curves) is remarkably increasing with the increase of the AGN contribution. In each panel we show the derived values of flux enclosed in the second broader Gaussian component, asymmetry $R$, prominence of the line wings $r9050$ and $w80$.}
\label{fig:gaussian_fits}}
\end{figure*}

\begin{figure}
\centering%
\includegraphics[angle=-90, width=\hsize ]{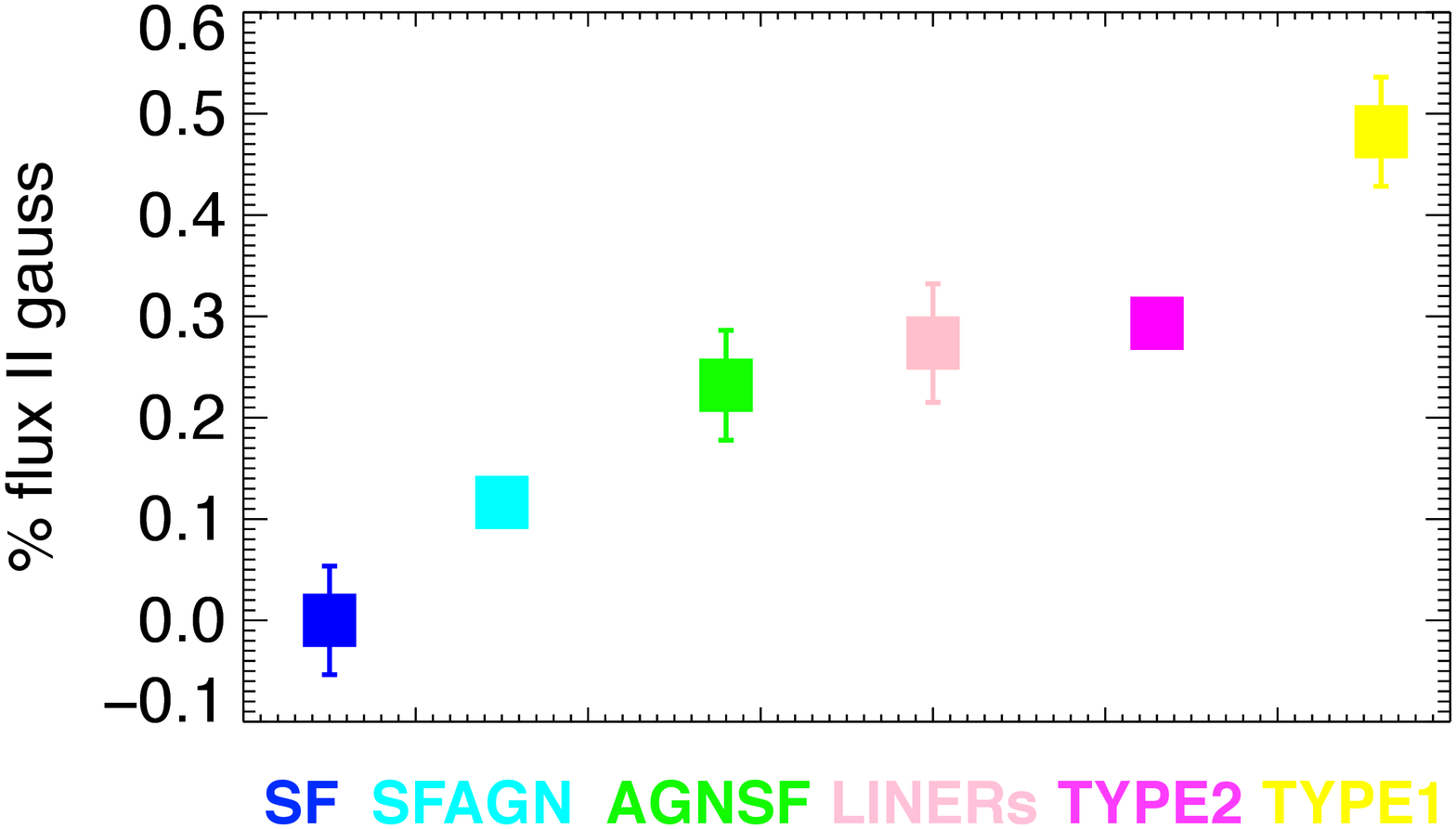}
\caption{Mean values of the flux percentage enclosed in the second Gaussian component for all the BPT classes: SF,SF-AGN, AGN-SF, LINERs, TYPE 2 and TYPE 1. The flux increasing with the increase of the AGN contribution. The error bars shown the dispersion around the mean values.}
\label{fig:flux_mean_bpt}
\end{figure}

{{Thus, we conclude that there is only marginal evidence for a broad component in the [OIII]$\lambda$5007 emission line profile and only in a specific locus of the SFR-stellar mass plane. When detected, such component is centered at the systemic redshift and there is no evidence of a blue-shift, as indication of wind, as in \cite{Mullaney2013}. Most of MS region is well represented by galaxies with [OIII] profile well fitted by a single Gaussian component with no asymmetry and with low kurtosis values. Only at high masses ($10^{10.5}-10^{11} M_{\odot}$) we observe a marginally higher kurtosis and so the presence of line wings. However, we do not find for these galaxies an excess of the line [OIII] width with respect to the galaxy velocity dispersion provided by the stellar component. Indeed, only a small percentage of the [OIII] flux is encapsulated in the wings. This indicates that likely a low percentage of the gas in these galaxies is moving away in a very low velocity wind.}}

\begin{figure*}
\centering
\includegraphics[angle=0 , width=0.33\textwidth]{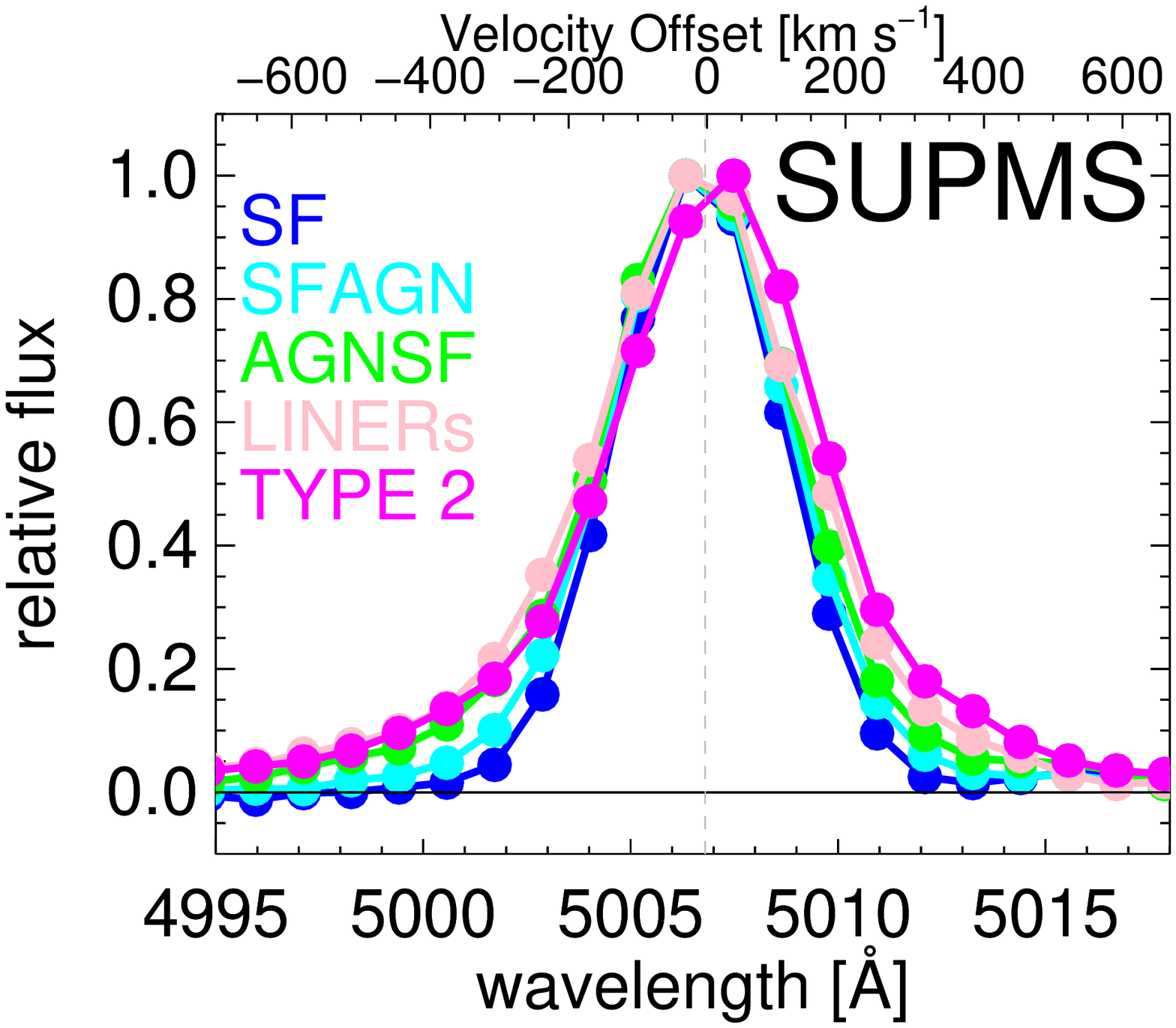}
\includegraphics[angle=0 , width=0.33\textwidth]{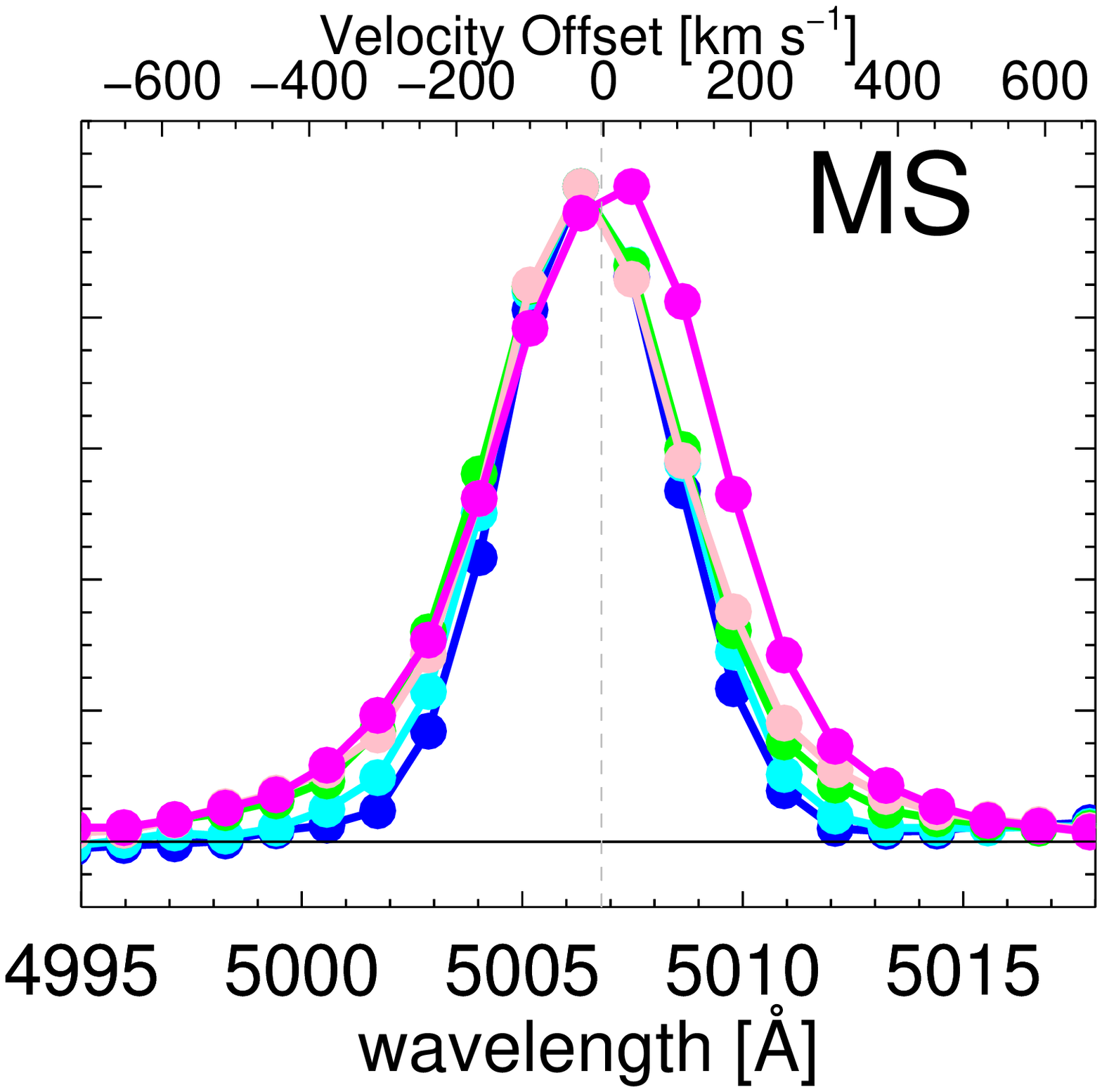}
\includegraphics[angle=0 , width=0.33\textwidth]{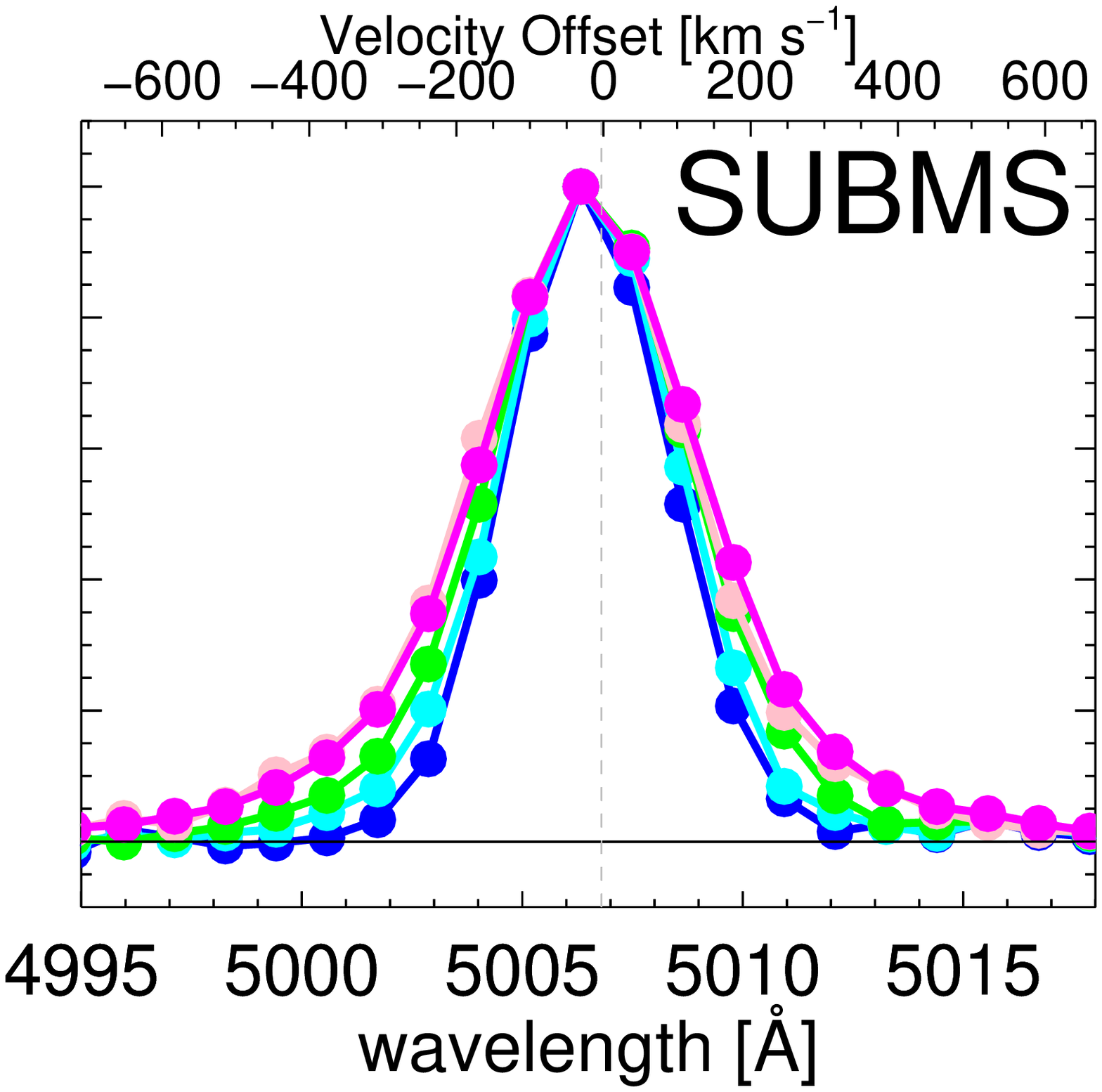}
\caption{Variation of the observed [OIII] emission line profile as a function of the SF end AGN contribution for galaxies located in different SFR bins. In the left, middle and right panels we show the galaxy bins located above (SUPMS), inside (MS) and below (SUBMS) the MS, respectively with $\Delta$SFR=$0.4$, $0.0$ and $-0.4$ dex. The galaxy bins showed have M$\star =10^{10.5}$ M$\sun$. The [OIII] emission line appear at any  M$\star$ and SFR}
\label{fig:SNR_bpt}
\end{figure*}   

\subsection{Trends with AGN and SF activity}
The absence of significant outflow signature in the global population does not exclude the possibility that strong winds might be observed in specific classes of objects. In order to investigate this possibility in this section we analyze the [OIII] line profile in the BPT subclasses, and so as a function of different ionization sources. As described in Section \ref{data} we split our sample in: SF, SF$-$AGN, AGN$-$SF, LINERs, TYPE 2 and unClass galaxies. We perform a first analysis on the stacked spectra of each subclass and, then, we study the stacked spectra as a function of the position in the SFR-M$_{\star}$ plane for each subclass separately.

Fig. \ref{fig:gaussian_fits} shows our multicomponent emission line fit to the stacked [OIII] line profile of each subclass. It is immediately apparent, that the significance of a second broad and blue-shifted component is remarkably increasing with the increase of the AGN contribution. While the star forming galaxies population show a symmetric Gaussian [OIII] line profile, the increase of the nuclear activity in the SF-AGN and AGN-SF leads to the raise of significant line wings. LINERs, TYPE 2 and TYPE 1 AGN, in particular, {show also a slight blue-shift of the broader component ( characterized by a velocity dispersion of $470.1 \pm 110.0$, $363.1\pm 14.0$ and $363.1\pm 14.0$ $km/s$, respectively) with respect to the systemic velocity ($\Delta V < 70 km/s$) and corresponding negative values in the asymmetry parameter.}
{{The flux enclosed in the second Gaussian component goes from 0\% in the SF galaxies 48\% in the TYPE 1 AGN, consistently with \cite{Mullaney2013}, as shown in Fig. \ref{fig:flux_mean_bpt}.}}
 
{The study of the [OIII] line profile of the individual BPT classes as a function of the location in the SFR-stellar mass plane shows the following aspects: 
\begin{itemize}
\item[-] SF galaxies, dominating the MS region up to masses of $10^{10.8} M_{\odot}$, are characterized by a purely Gaussian [OIII] line profile with no evidence of a second broader component. This is confirmed by the fitting procedure and by the non-parametric method that indicates values of asymmetry  and line wings consistent with the Gaussian values within $1.5\sigma$.
\item[-] SF galaxies with a small contribution from the central AGN (SF-AGN) show evidence of a second broader component only at the $2\sigma$ level. Such galaxies, as shown in Fig. \ref{pianoBPT}, are mainly located in the MS region at stellar masses $> 10^{10} M_{\odot}$.
\item[-] galaxies with a dominating AGN contribution (AGN-SF) show evidence of a second broader component at more the $3\sigma$ only on and above the MS. 
\item[-] LINERs and TYPE2 galaxies show a high SNR ($>3$) second broader component independently on their location on the SFR-stellar mass plane. See  Fig. \ref{pianoBPT} for their distribution in the plane.
\item[-] the results are confirmed by the non-parametric method. The analysis of the asymmetry R shows that for the SF object the line tend to be symmetrical (R$\sim0$) while in the LINERs and TYPE 2 we observe R$\sim -0.1$, consistent with the results of \cite{Zakamska2014} for a sample of SDSS obscured quasars.
\end{itemize}}

{An example of the AGN effect on the [OIII] line profile is shown in one mass bin above (left panel), on (central panel) and below (right panel) the MS in Fig. \ref{fig:SNR_bpt}.}

{The location of the AGN-SF, LINERs and TYPE2 AGNs perfectly matches the location of the plane where we observe in the global population a line wings value slightly larger than the Gaussian value. This is confirmed also by the BPT analysis applied to the stacked spectra, as a function of the location in the SFR-stellar mass plane. As shown in Fig. \ref{fig:plane_BPT_totalSample}, SF-AGN are preferentially located at high SFR and stellar masses, while AGN-SF and LINERs dominate the green valley region. Indeed, after removing such galaxies from the global sample, the value of the line wings parameter becomes consistent with the Gaussian value all over the plane. This indicates that the deviation from the pure Gaussian behavior observed in Fig. \ref{fig:r9050} is due to galaxies dominated by the AGN contribution. In turn, this suggests that, if the second broader component is interpreted as an indication of galactic wind, such wind is likely driven by the AGN, while SF seems not capable of driving any wind at any mass or SFR value.}

In order to better compare the [OIII] line width with the respect to the galaxy velocity dispersion in the different subclasses, we show in Fig. \ref{fig:sigma_oiii_sigma_star}  the [OIII] line width, measured by the standard deviation $\sigma_{[OIII]}$ as a function of stellar velocity dispersion $\sigma_{\star}$ for all the ionization classes: SF, SF$-$AGN, AGN$-$SF, LINERs and TYPE 2. Given that the median instrumental resolution of SDSS spectra is $\sim 70$ km s$^{-1}$ we restrict the analysis to the bins with $\sigma_{\star}$ above this limit. For clarity, we collect our $\sigma_{[OIII]}$ values in bins of $\sigma_{\star}$, with $\Delta \sigma_{\star}=20$ km s$^{-1}$. The different ionization sources are indicated with different colors as labeled in the figure. At fixed $\sigma_{\star}$, we find that the velocity dispersions measured for the ionized gas increases with the increase of the AGN activity from the "pure" SF to the TYPE 2 galaxies. 

\subsubsection{The unClass objects}
As mentioned in section \ref{bpt}, the unClass subsample includes a large amount of galaxies that are impossible to classify individually using the BPT diagram.
In order to take into account these large fraction of galaxies, we decide to perform the BPT analysis in the median stacked spectra. 
Following the approach of Concas et al. in prep. we use a combination of the publicly available codes pPXF \cite{Cappelalri_Emsellem2004} and GANDALF \cite{Sarzi2006} to fit and remove the stellar continuum and to derive emission line fluxes of the four emission lines used in the BPT diagram (i.e. H$\beta$, [OIII]$\lambda 5007$, [NII], H$\alpha$). 
As expected, the majority of the stacked spectra show very weak emission lines. These galaxies are mainly located in the so called quiescent region. At higher SFR values, all the unClass stacked spectra show all the four emission lines (H$\beta$, [OIII]$\lambda 5007$, [NII] and H$\alpha$) with  S/N$>4$ to classify them in the  BPT diagram. We check that these galaxies follow the same trend shown by the rest of the sample, with the prominence of the second broader component increasing in parallel to in increase of the nuclear activity contribution. 

\begin{figure}
\centering
\includegraphics[angle=-90 , width=\hsize ]{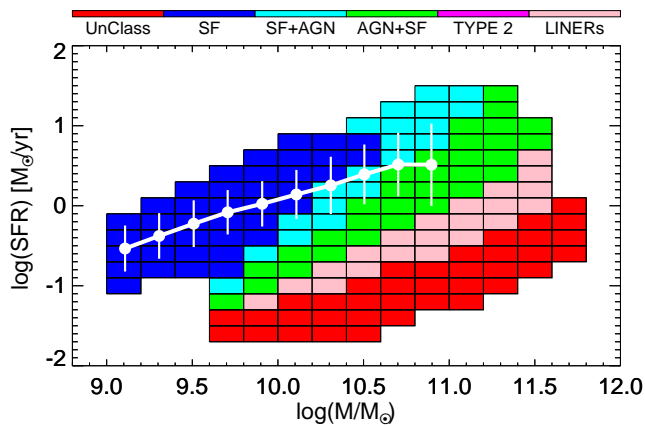}
\caption{BPT classification for the total sample in the SFR$-$M$\star$ plane.}
\label{fig:plane_BPT_totalSample}
\end{figure}

\begin{figure}
\centering
\includegraphics[angle=-90, width=1.12\hsize ]{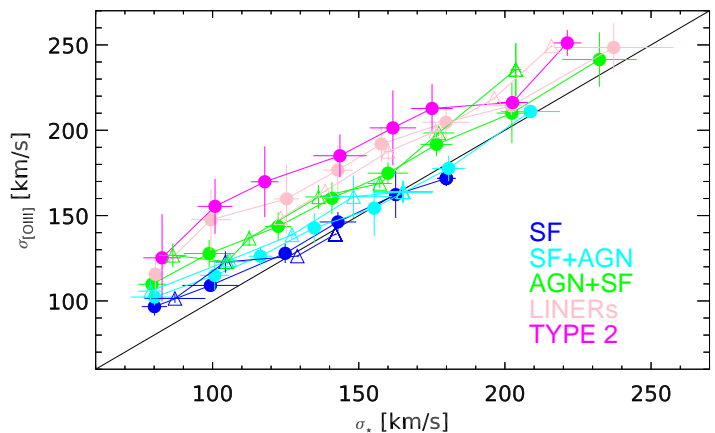}
\caption{$\sigma_{[OIII]}$ plotted against $\sigma_{\star}$ MPA-JHU mean values for the five ionization classes (SF, SF$-$AGN, AGN$-$SF, LINERs and TYPE 2 ). We show the median and dispersion values of $\sigma_{[OIII]}$ and $\sigma_{\star}$ in bins of $\sigma_{\star}$, with $\Delta \sigma_{\star}=20$ km s$^{-1}$. Filled symbols are the objects in the main SF, SF$-$AGN, AGN$-$SF, LINERs and TYPE 2 subsamples. Open triangles are the unClass objects. The solid black line denotes $\sigma_{[OIII]} = \sigma_{\star}$. }
\label{fig:sigma_oiii_sigma_star}
\end{figure}   


\section{Discussion and Conclusions}\label{conclusions}
In this work we investigate the presence of galactic winds in a large spectroscopic sample of $\sim 600000$ local galaxies drawn from the spectroscopic SDSS DR7 database. In particular, we use the deviation of the forbidden [OIII]$\lambda 5007$ emission line profile from a Gaussian as a proxy for the galactic winds. We use the spectral stacking technique to increase the SNR of the spectra and to determine how the average [OIII]$\lambda$5007 profile changes as a function of the key galaxy physical parameters, such as SFR and M$_{\star}$. We also explore how the line profiles relate to the particular photoionization mechanisms: SF or AGN activity.  We analyze the oxygen emission line profile by performing a line fit and a non-parametric analysis.
Our main results can be summarized as follows:
\begin{itemize}
\item {{ In the global galaxy population, we find no evidence of a second Gaussian broader component in most of the SFR-stellar mass plane. A marginal detection, at the $\sim 2\sigma$ level, is obtained only at stellar masses $>10^{10.5} M_{\odot}$ in a large range of SFR. This is confirmed by the observation of a line width parameter slightly larger (again at the $\sim 2\sigma$ level) than the value predicted for a pure Gaussian line profile in the same region of the plane.  The line profile appears to be always symmetric, even when a second broader component might contribute. The flux percentage enclosed in the broader component, when detected, is of the order to 10\% in most of the plane and it reaches values of 20-25\% at very high masses and SFR and in the green valley.}}

\item{{ The comparison of the line width of the [OIII] with the velocity dispersion obtained from the absorption stellar features reveals a good agreement in most of the plane, indicating that the [OIII] traces the underlying galaxy potential well as the stellar component. Only in few very low SFR and stellar mass bins we observed a disagreement, that we ascribe to spectral resolution issues and differences in the stellar and gas kinematics in purely disk galaxies.}}

\item{{The analysis of the  [OIII] line profile as a function of the BPT classification reveals that  for the "pure" SF galaxies, the ionized interstellar gas traced by the [OIII]$\lambda5007$ line never appears to be outflowing. The line profile is perfectly fitted by a single Gaussian without need of a second component. This holds in all the regions of the SFR-stellar mass plane dominated by SF galaxies, such as the MS. }}

\item{{The significance of a second broader Gaussian component increases with a clear trend with the increase of the AGN contribution to the galaxy spectrum, with a maximum for the AGN TYPE1 of \cite{Mullaney2013}. The flux enclosed in the second component rises steadily from 0\% in pure SF galaxy to $\sim$48\% in the TYPE1 AGNs.}} 

\item{{The analysis of the [OIII] line profile of each BPT class in the SFR-stellar mass diagram shows that galaxies with an increasing AGN contribution occupy preferentially the region of the diagram where the global population show a marginal deviation from the Gaussian line profile: at high mass and SFR and in the green valley. If AGN hosts are removed from the global sample, such deviations disappears in any locus of the SFR-stellar mass digram. The preferential location of AGN hosts, distinguished in different BPT classes, is also analysed by performing the BPT analysis on the stacked spectra. We clearly observe a preferential location of galaxies in the diagram as a function of their nuclear activity. In particular, SF-AGN galaxies occupy the region at high stellar mass and SFR, while AGN-SF and LINERs are located preferentially in the green valley.}}

\item{{The comparison of the velocity dispersion obtained from the [OIII] line width with the velocity dispersion derived from the absorption features imprinted by the stellar component, shows that, the ionized gas traces the galaxy potential well as the stellar component for the pure SF galaxies. The ratio between the two velocity dispersions increases at the increase of the AGN contribution at any bin of stellar velocity dispersion. This, once again, confirms the role of the AGN contribution in leading to a larger [OIII] line width, and so to the presence of winds, independently on the galaxy mass and SFR.}}

\end{itemize}


\begin{figure}
\centering
\includegraphics[angle=-90 , width=\hsize ]{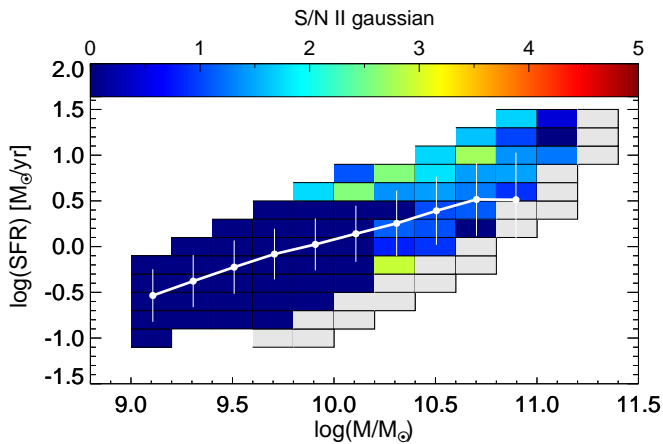}
\caption{Signal-to-noise ratio, SNR, of the second broader Gaussian component for the "HII" galaxy sample in the SFR-M$_{\star}$. The white line shows the mode and dispersion of the MS. At high SFR and M$\star$ the SNR shows a slight increase. The galaxy bins whit total emission line SNR $< 8$ are plotted in grey color.}
\label{fig:fluxHII}
\end{figure}   


\begin{figure*}
\centering
\includegraphics[angle=-90 , width=0.8\textwidth]{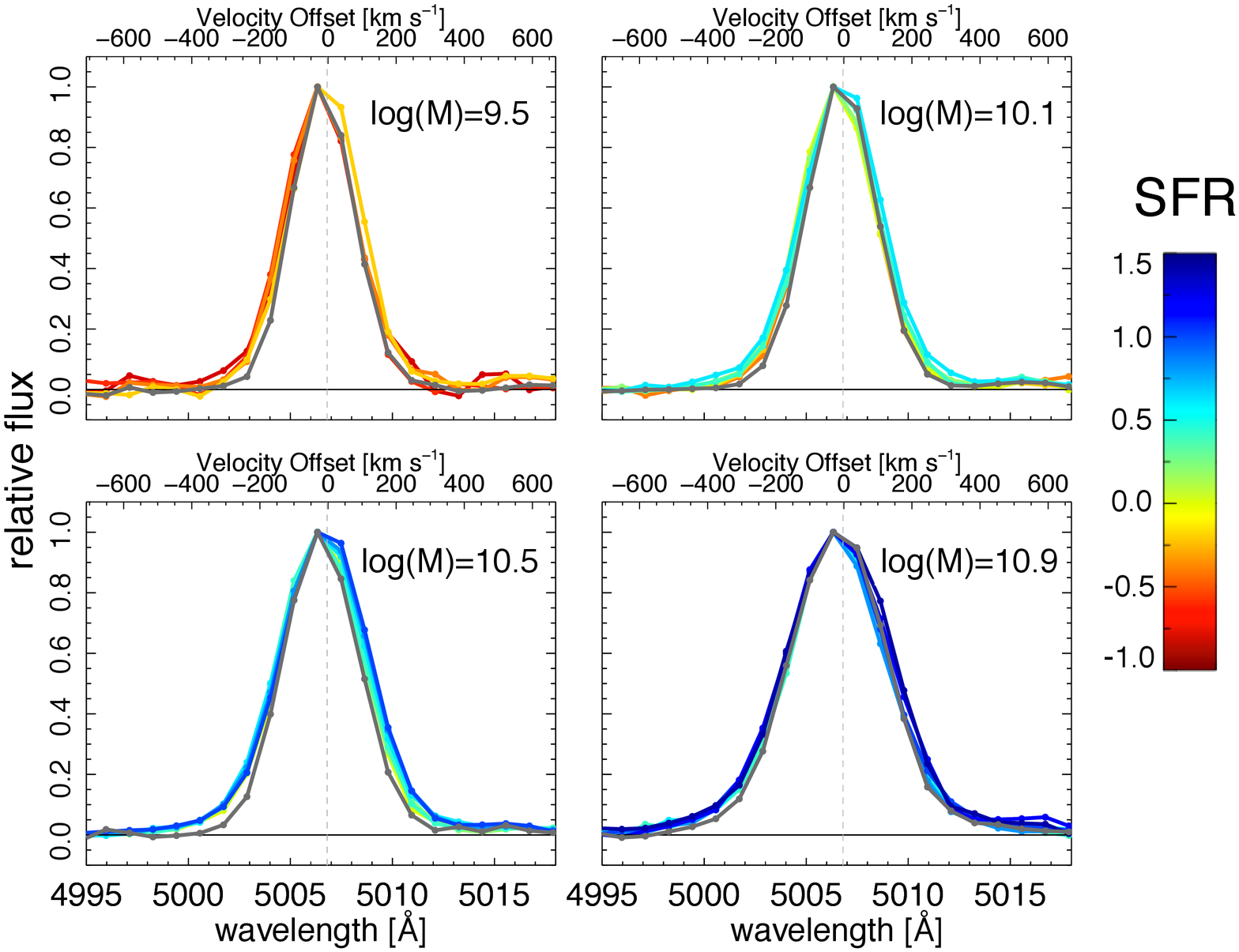}
\caption{Comparison in four different stellar mass bins of the observed [OIII] line profile of pure SF and SF-AGN galaxies, respectively, in the HII sample, classified according to the BPT classification of \cite{Stasinska2006}.  The line profile of the pure SF galaxies does not change as a function of the SF level and it is indicated by the grey curve in each panel. The line profile of the SF-AGN galaxies tends to increase in width at increasing SFR as indicated by the curves color-coded as a function of the SFR.}
\label{nuova_immagine}
\end{figure*}   


{{Our results clearly indicate that, when using the low density ionized gas emission lines as wind indicator,  the SF activity itself, in the local Universe, is not capable of driving galactic winds at any value of the instantaneous SF rate or stellar mass. We point out that this outcome is not directly in disagreement with previous findings of outflow signatures in SF galaxies with much higher level of SF activity, as ULIRGS or high redshift Main Sequence star forming galaxies \citep[e.g.][]{Heckman+90,Pettini2000,Shapley2003,Rupke2002,Rupke2005a,Rupke2005b,Martin2005,Martin2006, HillZakamska2014}. Indeed, such systems exhibit SFR level at least one order of magnitude higher than the level of activity in the local average star forming galaxies. Thus, we can not exclude that star formation might be able to drive outflow when the feedback are sufficiently energetic to induce motion. Our results indicate, although, that, at the level of activity observed in the local Universe, SF feedback is not capable of inducing significant gas bulk motions. A direct comparison with a similar sample of galaxies at higher redshift, and so with an higher level of SF activity, must be carried out to understand which is the SF threshold, above which SF might eventually lead to systematic galactic winds.}}

{{
This result, however, contrasts with the findings of \cite{Cicone2016}, who detect signature of ionized outflows in local SDSS star forming galaxies at high value of SFR. They find also an increase of the line width at increasing SFR at fixed mass, and interpret this as evidence of SF driven outflow. We ascribe such discrepancy to the differences in the sample selection. \cite{Cicone2016} analyze a sample of $\sim 160000 $ local star-forming galaxies classified in the BPT diagram according to \cite{Kauffmann2003b}. 
As showed in \cite{Stasinska2006}, this selection criteria may introduce a significant bias in the SF galaxy sample, due to the inclusion of a large percentage of systems with an AGN contribution. This effect may be particularly strong at high SFRs where the number of composite SF$-$AGN is high. In order to investigate this issue, we perform our analysis on the ``HII'' sample of star forming galaxies as defined in \cite{Cicone2016}. As \cite{Cicone2016}, we do not find any significant broad second Gaussian component for the galaxies located at low SFR and M$_{\star}$, while we note a weak increase of this second component ($\sim 10 \%$ of the total line flux, $\sim$2 $\sigma$ level) at SFR$\geq 1 M_{\sun}/yr$ and stellar masses M$_{\star} \geq 10^{10} M_{\sun}$ ( Fig. \ref{fig:fluxHII}). However, by classifying the \cite{Cicone2016} ``HII'' sample with the \cite{Stasinska2006} BPT classification, as done for our sample, we see (Fig. \ref{nuova_immagine}) that
\begin{itemize}
\item pure SF galaxies show always a pure Gaussian profile with constant width at fixed stellar mass;
\item SF$-$AGN galaxies show a more prominent second Gaussian component, although of low significance ($< 2\sigma$), a larger width with respect to pure SF galaxies of the same mass and SFR, and an increase of the width as the SFR increases at fixed stellar mass.
\end{itemize}
We, thus, conclude that the effect observed by \cite{Cicone2016} is likely due to the contamination by systems with an AGN contribution. The larger width of the [OIII] profile at high SFR in such galaxies could be due to availability of a large quantity of cold gas, which favors at the same time the SF activity and the accretion onto the central black hole, eventually leading to stronger BH feedback.
We point out also that the transition between pure SF galaxies and AGN contaminated galaxies takes place in the region of the diagram that start to be dominated by bulgier sources (see central panel of Fig. \ref{fig:planeBsuTr}). This is a further indication that the central activity might be a bias in the analysis performed by \cite{Cicone2016}. }}

{{Our results confirm the role of AGN feedback in leading to galactic wind, at least of ionized gas, also in the local Universe. This result is in agreement with many recent findings in a large redshift window. At low redshift, we find very good agreement with the results of \cite{Mullaney2013}, that observe a prominent blue-shifted and broader Gaussian component in addition to the systemic one for TYPE 1 and TYPE 2 AGNs. Our analysis extend this result also at lower level of nuclear activity. While the use of the fiber SDSS spectra does not allow us to have any spatial information on the origin and the location of the wind, by using the optical integral field unit (IFU) observations of sixteen TYPE 2 AGN selected from the \cite{Mullaney2013} parent sample, \cite{Harrison2014} demonstrate that this particular [OIII] emission line features is due to the kiloparsec scales outflows, extended over $\geq 6-16$ kpc.

At higher redshift, \cite{Brusa2015} and \cite{Cresci2015} show that in X-ray bright AGN, the nuclear activity can lead to very powerful winds, of the order of $\sim 1000$ km/s, also in the molecular gas.  

We point out that, on average, we do not observe such large velocity shift or line width. The observed average velocity shift with respect to the systemic redshift is below the SDSS instrumental resolution $\Delta V < 70 km/s$, and the line width of the second Gaussian component is only of the order of 350-400 $km/s$ also in the BPT classes, which include an AGN contribution. Thus, rather than galactic wind, the overall population of AGN hosts in the local Universe undergo a phase of light {\it{``breeze''}}. This difference might arise for two reasons: a) bright X-ray AGNs are rare objects in the local Universe, thus, the average velocity shift and line width is dominated by sources with a lower level of nuclear activity and feedback; b) we are observing a final phase of the wind, which was more powerful in the past due to the higher nuclear activity of the central AGN.

Since the AGN hosts are located in the high mass region of the SFR-stellar mass diagram at stellar masses M$_{\star} \geq 10^{10.5} M_{\sun}$, thus likely in dark matter halos larger than $10^{12-12.5} M_{\sun}$, such velocity are at least one order of magnitude lower than the escape velocity from the galaxy. So the gas entrained in the wind is very likely not able to escape the galaxy potential well. 

Thus, we conclude that, at least in the local Universe, the AGN activity is likely the only mechanism capable of driving galactic winds. However, given the velocity of such wind, this is not able to let the gas escape from the galaxy and so affect the galaxy gas content and SF activity.
}}

\begin{acknowledgements}
 The authors are grateful to the anonymous referee, whose suggestions improved this manuscript. MB thanks the DFG cluster of excellence 'Origin and Structure of the Universe' (www.universe-cluster.de) for partial support during the completion of this work, and acknowledges support from the FP7 Career Integration Grant "aEASy" (CIG 321913).

\end{acknowledgements}

%
%

%
%

\bibliographystyle{aa}
\bibliography{alice} 

\end{document}